\documentclass[twocolumn, apj, iop, numberedappendix, appendixfloats, twocolappendix, tighten]{openjournal}

\usepackage[T1]{fontenc}

\DeclareRobustCommand{\VAN}[3]{#2}
\let\VANthebibliography\thebibliography
\def\thebibliography{\DeclareRobustCommand{\VAN}[3]{##3}\VANthebibliography}

\usepackage{graphicx}	
\usepackage{amsmath}	
\usepackage{amssymb}	

\usepackage{xcolor}
\usepackage{hyperref}
\hypersetup{
    colorlinks=true,
    linkcolor=blue,
    citecolor=blue,
    urlcolor=blue
}
\newcommand{\orc}{\includegraphics[height=\fontcharht\font`A]{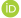}}
\newcommand{\orcid}[1]{\href{https://orcid.org/#1}{\orc}}

\usepackage[normalem]{ulem}

\usepackage{indentfirst}

\usepackage{newtxtext,newtxmath}

\usepackage{threeparttable}

\graphicspath{{./}{figures/}}

\def\gsim{\;\rlap{\lower 2.5pt
 \hbox{$\sim$}}\raise 1.5pt\hbox{$>$}\;}
\def\lsim{\;\rlap{\lower 2.5pt
   \hbox{$\sim$}}\raise 1.5pt\hbox{$<$}\;}
\newcommand{\Da}{Damk{\"o}hler }

\usepackage{orcidlink}

\begin{document}

\title{From Clumps to Sheets: Geometry Controls the Temperature PDF of Multi-Phase Gas}

\author{Zirui Chen\orcidlink{0000-0001-8755-3836}}
\email{ziruichen@ucsb.edu}

\author{S. Peng Oh\orcidlink{0000-0002-1013-4657}}
\affiliation{
Department of Physics,
University of California, Santa Barbara,
Santa Barbara, CA 93106, USA
}

\begin{abstract}
Temperature probability distribution functions (PDFs) are a compact description of the thermal structure of multi-phase turbulent gas, and are directly linked to observables such as emission/absorption line ratios and phase mass fractions. In the circumgalactic medium (CGM) literature, temperature PDFs are often interpreted using planar turbulent radiative mixing layers, for which analytic models successfully reproduce the simulated temperature structure. These PDFs are assumed to be universal. By contrast, studies of the multiphase interstellar medium (ISM) typically use turbulent-box simulations, which produce broad PDFs but lack a clear theoretical interpretation. Using 3D hydrodynamic simulations under both ISM and CGM conditions, we compare planar mixing layers with turbulent-box simulations under identical microphysical conditions. Despite identical cooling and turbulent driving, the resulting temperature PDFs differ substantially. The missing ingredient is geometry. We demonstrate that the temperature PDF can be decomposed into the product of the area of temperature isosurfaces and the thickness of the corresponding temperature layers. The thickness is controlled primarily by microphysics, such as radiative cooling and thermal conduction, and is well captured by existing mixing-layer models. The isosurface area, however, is set by morphology. In mixing layers it remains sheet-like, whereas in turbulent media cold gas forms clumps whose interfaces expand with temperature and eventually percolate into connected sheets. This geometric transition produces broad PDFs with large intermediate-temperature mass fractions. These results have implications for long-standing puzzles such as thermally unstable gas in the ISM, the large O{\footnotesize VI} reservoir in the CGM, and X-ray-H$\alpha$ correlations in jellyfish tails.
\end{abstract}


\maketitle



\section{Introduction} \label{sec:intro}

Gas in and around galaxies is highly multiphase: radiative cooling and heating create thermally stable phases whose densities and temperatures differ by orders of magnitude, while turbulence continuously mixes these phases and populates intermediate temperatures. A natural way to characterize this thermal structure is through the temperature probability density function (PDF), which describes how much gas occupies each temperature interval. Temperature PDFs are directly accessible in simulations and strongly affect observables such as phase mass fractions and emission or absorption line ratios. In quasi-isobaric environments, they provide a compact description of the thermal phase structure of the gas. 

The temperature distribution of multiphase gas is central to several astrophysical environments. In the ISM, gas spans many orders of magnitude in temperature and density while remaining in rough pressure equilibrium \citep{mckee77, cox05}; the neutral medium is dominated by the CNM and WNM at $T \sim 50$ K and $T \sim 7000$ K, respectively \citep{Wolfire1995, Draine:2011, murray18, McClure-Griffiths2023}, and 21-SPONGE observations suggest that a substantial fraction of H~I resides at thermally unstable temperatures \citep{murray15-SPONGE, murray18, koley19}, presumably because turbulence mixes gas across phase boundaries \citep{elmegreen04, heiles03, Saury:2014}. In the CGM, a hot volume-filling phase at $T \sim 10^6$ K coexists with cool cloudlets at $T \sim 10^4$ K \citep{tumlinson17, FGOH23}; shear-driven turbulence generates radiative mixing layers that are central to models of cold-gas survival and intermediate-temperature emission and absorption, and may help explain the large O\textsc{vi} reservoir inferred from \textit{COS} observations \citep{Tumlinson2013, werk14, prochaska17, mcquinn18}. Similar issues arise in jellyfish tails, cluster filaments, and stellar or AGN outflows \citep{moretti18, sun22, sparks09, donahue00}: in all cases, the abundance of intermediate-temperature gas is both physically informative and observationally consequential.

A common approach towards tackling the problem of multi-phase turbulence involves running controlled 3D hydrodynamic simulations. One classic simulation setup zooms into the interface between the hot and cold gas phases and studies the plane-parallel turbulent radiative mixing layers that form due to Kelvin-Helmholtz instability \citep{kwak10,ji19, fielding20, tan21, zhao23,sharma25}. The assumption is that mixing layer widths are much smaller than other lengthscales such as clump size, so the restriction to planar geometry is valid. A key insight from these simulations is that the relative strength of cooling, heating, and turbulence in a multiphase turbulent medium can be captured by a single dimensionless parameter: the \Da number \citep{damkohler40,tan21}, 
\begin{align}
    {\rm Da} = \frac{t_{\rm mix}}{t_{\rm cool}(T_{\rm mix})}. 
    \label{eq:damkohler number}
\end{align} 
Here, $t_{\rm mix} = \left. L \right/ v_{\rm turb}$ is the mixing or eddy turnover timescale, and $t_{\rm cool}(T_{\rm mix})$ is a characteristic cooling time typically evaluated at the geometric mean temperature of the thermally stable phases participating in the mixing \citep{begelman90,gronke18}. Mixing layer simulations show that the \Da number divides a turbulent medium into two regimes \citep{tan21}. When ${\rm Da} \gg 1$, net cooling wins over turbulent mixing and gas settles into the thermally stable phases connected by a thin, grid-scale interface. When ${\rm Da} \lesssim 1$, turbulent mixing dominates and produces abundant intermediate temperature, thermally unstable gas, resulting in a thick, blurry interface region. Cold gas grows in the former case but gets destroyed via mixing in the latter case. Turbulent mixing layer simulations have been helpful in understanding cold gas survival in the CGM \citep{gronke18, gronke20-cloud, chen23_molecules_MNRAS} and predicting emission features originating from intermediate temperature gas residing in the mixing layers \citep{tan21-lines, chen23, chenpeng26}. 

Another prominent type of numerical simulation considers a representative patch in the ISM or CGM and analyzes the dynamics and phase structure of gas under driven turbulence. In the context of the ISM, such simulations have been used to study the phase distribution of multiphase turbulent gas \citep{piontek05, audit05, Saury:2014}, 
as well as its turbulent cascade \citep{Beattie25,Connor26}. Similar turbulent box simulations are also studied under CGM conditions for cold gas survival \citep{gronke22-turb,das23} and the correlation between cold and hot phase velocity structure functions \citep{mohapatra22-turb_box}. 

These two approaches emphasize different aspects of the same physics. Mixing-layer models capture the microphysics of interface structure --- in particular the roles of cooling, conduction, and enthalpy transport --- and have been especially powerful in the CGM context. Turbulent-box simulations, by contrast, capture the global morphology of cold gas embedded in a turbulent medium. Whether temperature PDFs calibrated on planar mixing layers apply to such more complex turbulent environments remains uncertain. Using controlled 3D hydrodynamic simulations, we compare planar turbulent mixing layers with turbulent-box simulations under otherwise identical microphysical conditions. Despite identical cooling, conduction, and turbulent driving, the resulting temperature PDFs differ dramatically. 

In this work we show that the missing ingredient is geometry. The temperature PDF can be understood geometrically by decomposing the volume of gas at each temperature into the product of the area of the corresponding temperature isosurface and the thickness of that temperature layer. The thickness is governed primarily by microphysics and is well captured by existing mixing-layer models; the isosurface area depends on the morphology of the cold gas structures. As a result, geometry can dominate the temperature PDF and provides a natural framework for interpreting both mixing-layer and turbulent-box simulations.

The outline of this paper is as follows. In \autoref{sec:methods}, we describe the setup of our turbulent box and turbulent mixing layer simulations, as well as implementation of radiative cooling and thermal conduction. In \autoref{sec:results}, we present novel results from our simulations, including how turbulent box and mixing layer simulations under the exact same conditions show distinct temperature PDFs and how turbulent box PDFs are sensitive to the turbulent Mach number ($\mathcal{M}_{\rm turb}$) and the \Da number (${\rm Da} = \left. t_{\rm mix} \right/t_{\rm cool}$). In \autoref{sec:interpretation}, we explain these results by performing a geometric decomposition of our temperature PDFs into temperature isosurface area times the thickness of isosurfaces, which allows us to understand the morphology of intermediate temperature interfaces. Finally, in \autoref{sec:summary} we discuss the implications of our work and summarize the main conclusions.

\section{Methods} \label{sec:methods}

We seek to use 3D hydrodynamic simulations to understand what physical processes set the temperature PDF of a multiphase turbulent medium. Our simulations are performed using Athena++ \citep{stone20}, which solves the 3D hydrodynamic equations on a uniform Cartesian grid using the HLLC Riemann solver. We primarily employ two simulation setups: turbulent boxes and plane-parallel turbulent mixing layers. In addition, we run two wind tunnel simulations 
to directly compare PDFs of wind tunnel simulations and mixing layer simulations. In this section, we describe all three simulation setups as well as how we implement important physical processes such as radiative cooling and conduction. We provide a summary of all the simulations we run in \autoref{tab:summary of simulations}. In \autoref{sec:convergence test}, we perform a convergence test to demonstrate that our simulation results are not affected by resolution.

\subsection{Cooling and Heating Functions} \label{sec:radiative cooling}

\begin{figure}
\centering
\includegraphics[width=\columnwidth]{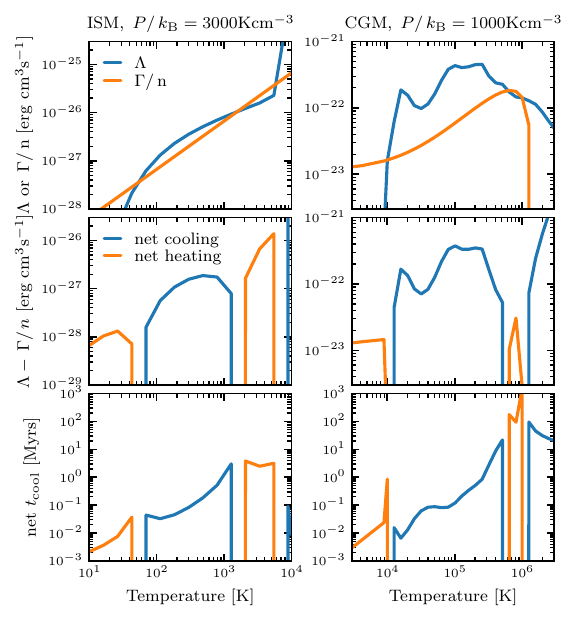}
\caption{Cooling ($\Lambda$) and Heating ($\left. \Lambda\right/ n$) functions (\textit{top}), net cooling functions ($\Lambda - \left. \Lambda\right/ n$, \textit{middle}), and net cooling time profiles (\textit{bottom}) we use for ISM (\textit{left}, following \protect\cite{Koyama:2002}) and CGM (\textit{right}, following \protect\cite{gnat07}) conditions. For ISM conditions, we choose $\left. P \right/k_{\rm B}=3000 {\rm K} {\rm cm}^{-3}$, which yields stable thermal equilibrium temperatures at $\sim 60$K and $\sim 6000$K. For CGM conditions, we choose $\left. P \right/k_{\rm B}=1000 {\rm K} {\rm cm}^{-3}$ and use a toy density dependent heating rate such that the net cooling function is thermally bistable at $10^4$K and $10^6$K. This is identical to the setup in \protect\cite{tan21}. By design, the net cooling function and the net cooling time profile for ISM and CGM are qualitatively similar.}
\label{fig:net_cooling_curves_ISM_CGM}
\end{figure}

When implementing cooling and heating, we utilize the module developed in \cite{tan23-cloud-atlas}, which implements the net cooling function: 
\begin{align}
    \rho \mathcal{L} = n^2 \Lambda - n \Gamma,
    \label{eq:net cooling}
\end{align} 
where $\Lambda$ is the cooling function, and $\Gamma$ is the heating rate, using the  fast and robust exact cooling algorithm described in \cite{townsend09}. In the absence of other physics, this drives the gas to thermal equilibrium temperatures where net cooling vanishes. For the majority of this work, we use the cooling and heating functions appropriate for ISM conditions. However, in \autoref{sec:CGM-extension}, we show simulation results using the CGM net cooling function to demonstrate the general applicability of our conclusions to a wide range of astrophysical scenarios.

In what follows, we describe the net cooling functions for both ISM and CGM conditions. The fiducial choices of pressure we use are $\left. P \right/k_{\rm B}=3000 {\rm K} \, {\rm cm}^{-3}$ for ISM conditions and $\left. P \right/k_{\rm B}=1000 {\rm K} \, {\rm cm}^{-3}$ for CGM conditions. These fiducial choices yield the cooling and heating functions shown in \autoref{fig:net_cooling_curves_ISM_CGM}.

\subsubsection{Net Cooling Function in the Interstellar Medium} \label{sec:ISM cooling}

Under ISM conditions, the cooling ($\Lambda$) and heating ($\Gamma$) functions are described by \cite{Koyama:2002}. Within a pressure range $P_{\rm min} < P < P_{\rm max}$, the thermal equilibrium condition $\mathcal{L} = 0$ is met at 3 temperatures: two thermally stable phases with an intermediate, thermally unstable phase. This thermally bistable phase structure has been studied extensively in previous works \citep[e.g.][]{mckee77-ISM, jennings21}.

In the left column of \autoref{fig:net_cooling_curves_ISM_CGM}, we show these cooling and heating functions in the ISM temperature range of $\sim 10-10^4$K. Under isobaric conditions, the net cooling time $t_{\rm cool}$ is: 

\begin{align}
    t_{\rm cool}=\frac{\mu_{\rm e} \mu_{\rm H}}{\left(\gamma - 1\right) \mu^2} \frac{k_{\rm B}^2 T^2}{P \Lambda_{(T)}},
    \label{eq:net t_cool}
\end{align} 
where $\mu = \left. \rho \right/n$, $\mu_{\rm e} = \left. \rho \right/n_{\rm e}$, $\mu_{\rm H} = \left. \rho \right/n_{\rm H}$, and $t_{\rm cool}$ in the bottom left panel of \autoref{fig:net_cooling_curves_ISM_CGM} is evaluated using $\left. P \right/k_{\rm B}=3000 {\rm K} \, {\rm cm}^{-3}$. Note that starting from $T \sim 70$K, the net cooling time profile steeply increases and reaches its maximum value at $\sim 3000$K.

\subsubsection{Net Cooling Function in the Circum-galactic Medium} \label{sec:CGM cooling}

The CGM is multiphase \citep{tumlinson17, FGOH23}, with cold clouds at $T\sim 10^4$K embedded in a hot, volume-filling phase at the virial temperature, which is at $T\sim 10^6$K for Milky Way like galaxies. In this temperature range,  we adopt the collisional ionization 
equilibrium (CIE) cooling curve in \cite{gnat07} and assume solar metallicity (X = 0.7, Z = 0.02). We note that although CIE is a good approximation for the low density hot phase of the CGM, it can fail at lower temperatures where the cooling time can fall below the recombination time \citep{tumlinson17, Oppenheimer13}, or when photoionization changes the abundance of key coolants \citep{slavin93,ji19}. Additionally, the metallicity of the CGM is not strictly solar and can vary by over an order of magnitude across outflows, recycled gas, and different radial distances from central galaxy \citep{Lehner13, prochaska17}. These caveats can affect the exact shape of the cooling function we adopt. However, as we will demonstrate later in this work, our main conclusions are not sensitive to the exact shape of the cooling function and in fact hold across both ISM and CGM conditions.

Unlike the ISM, the heating function in the CGM is not well known. We use a toy density dependent heating rate such that the net cooling function is thermally bistable at $10^4$K and $10^6$K, with a thermally unstable equilibrium at $6 \times 10^5$K\footnote{We also run a couple of simulations under ICM conditions, with $T_{\rm hot} = 10^7$K (see \autoref{tab:summary of simulations} for details). For those simulations, we adjust the density dependent heating rate such that the net cooling function is thermally bistable at $10^4$K and $10^7$K, with a thermally unstable equilibrium at $6 \times 10^6$K.}. This ensures numerical stability and is identical to the setup in \cite{tan21}. The cooling ($\Lambda$) and heating($\left. \Gamma \right/n$) functions, net cooling function ($\Lambda - \left. \Gamma \right/n$), and net cooling time ($t_{\rm cool}$) we use for the CGM are shown in the right column of \autoref{fig:net_cooling_curves_ISM_CGM}, where we have used a typical CGM pressure of $\left. P \right/k_{\rm B}=1000 {\rm K} \, {\rm cm}^{-3}$ when evaluating the net cooling time.

\subsection{Thermal Conduction} \label{sec:thermal conduction, numerical implementation}

The conductive heat flux is given by $Q = -\kappa \nabla T$, where $\kappa$ is the conductivity. Without explicit treatment of thermal conduction, our simulations only rely on numerical diffusion, which is equivalent\footnote{We can understand this from the fact that advection errors cause numerical diffusion of thermal energy, with ${\mathbf F} = - \rho c_{\rm P} \alpha \nabla T$, where $c_{\rm P} = 5/2 (k_{\rm B}/\mu m_{\rm H})$ and $\alpha \sim v \Delta x$, where $v$ is a characteristic velocity (here, the shearing velocity) and $\Delta x$ is the spatial resolution \citep{robertson10}. Setting ${\mathbf F} = - \kappa \nabla T$ gives $\kappa \propto \rho \propto T^{-1}$, under isobaric conditions, which is verified in numerical simulations \citep{tan21-lines}.} to $\kappa \propto T^{-1}$. However, it is useful to consider some other forms of conductivity appropriate to the ISM and the CGM. In the ISM where neutral atoms are dominant, conductivity is given by \cite{parker53}:

\begin{align}
    \kappa_{\rm ISM} = 2.5 \times 10^3 T^{0.5} {\rm cm}^{-1} {\rm K}^{-1} {\rm s}^{-1}.
    \label{eq:kappa ISM}
\end{align} 
For an ionized plasma like the CGM, conductivity is given by \cite{spitzer62}:

\begin{align}
    \kappa_{\rm sp} = 5.7 \times 10^{-7} T^{2.5} {\rm erg} \, {\rm cm}^{-1} {\rm K}^{-1} {\rm s}^{-1}.
    \label{eq:kappa ISM}
\end{align}

\cite{tan21-lines} showed that in the CGM regime, temperature PDF of turbulent mixing layers are sensitive to the temperature dependence but \textit{not} the normalization of the conductivity $\kappa$. We will expand upon this and explore whether these conclusions hold in the ISM regime. To do so, we employ a two moment approximation method for thermal conduction. This approach is described in detail in \cite{jiang18}, where it is used for implementing cosmic rays. We closely follow \cite{tan21}, where this approach was used to implement thermal conduction.\footnote{We refer the readers to Section 2.2 of \cite{tan21} for a more detailed description of the numerical details.} 

\subsection{Turbulent Box Simulation Setup} \label{sec:turbulent box sim setup}

The initial condition of our turbulent box setup consists of a cold cloud at the cold stable equilibrium temperature of the net cooling function ($\sim 60$K for the ISM net cooling function and $10^4$K for the CGM net cooling function) embedded in a hot background at the hot stable equilibrium temperature ($\sim 6000$K for ISM and $10^6$K for CGM). To avoid the carbuncle instability, which seeds grid-aligned numerical artifacts \citep{Moschetta2001}, the initial cold cloud is not spherical but rather potato-shaped.\footnote{We refer the readers to \cite{gronke22-turb} for a visualization of this initial condition.} Additionally, we introduce percent-level density fluctuations to the initial cloud. The initial cloud size is chosen such that it satisfies the survival criterion in a turbulent medium \citep{gronke22-turb}. For example, in a turbulent medium with density contrast $\chi \sim 100$ and $\mathcal{M}_{\rm turb} = \left.  v_{\rm turb}\right/ c_{\rm s,hot} \sim 1$, the survival criterion is $r_{\rm cloud} \gtrsim 200 l_{\rm shatter}$, where $l_{\rm shatter}$ is the characteristic scale of cooling induced fragmentation given by $l_{\rm shatter} \sim c_{\rm s,cold}$ $t_{\rm cool,cold}$ \citep{mccourt18, gronke20-mist}. When implementing the ISM net cooling function, we also experimented with another initial condition: a uniform box at the unstable equilibrium temperature ($\sim 1000$K) with percent level density fluctuations. This initial condition is more analogous to existing numerical studies of thermal instability \citep{kritsuk02, vazquez_semadeni_03, piontek04, audit05}. We found that it yields identical results as our fiducial cold cloud in hot background initial condition. 

We prescribe periodic boundary conditions in all three spatial dimensions and continuously drive turbulence in the box on the largest scales\footnote{We stir at scales $k=\left. 2\pi n \right/ L_{\rm box}$, with $0 < n < 2$. This is similar to the setup in \cite{gronke22-turb}.} from the start of the simulation. The driven turbulence follows a Kolmogorov spectrum with $f_{\rm sol} = \left. 1 \right/ 2$ (i.e., 50\% solenoidal and 50\% compressive driving) and cascades down to small scales as the simulation progresses. With the driven turbulence described above, the turbulent velocity $v_{\rm turb}$ in the box quickly settles into a time-steady value after about an eddy turnover time $t_{\rm mix} = \left. L \right/ v_{\rm turb}$, where L is the box size. We have the freedom to change the value of $v_{\rm turb}$ by changing the energy injection rate of the turbulence driver. Additionally, we implement the relevant net cooling functions as described in \autoref{sec:radiative cooling}.

We vary the \Da number (\autoref{eq:damkohler number}) by varying the strength of turbulent driving, thereby changing $v_{\rm turb}$ and hence $t_{\rm mix}$. Note that this choice changes two dimensionless parameters, Da and the Mach number $\mathcal{M}$. It would be cleaner to instead vary $t_{\rm cool}$, as in previous simulation setups \citep{tan21}.  However, at fixed metallicity, $t_{\rm cool}$ can only change due to a change in gas pressure. While a large dynamic range in gas pressure is realistic in the CGM, in the ISM, thermally bistable gas only exists in a narrow range of pressures, so it is unrealistic to artificially vary $t_{\rm cool}$; varying $t_{\rm mix}$ more faithfully represents what happens in the real ISM. Thus, it is the standard choice in ISM simulations (e.g., \citealt{Saury:2014}). 

In a typical turbulent box simulation, the cold gas mass grows until eventually the temperature PDF becomes time-independent. We terminate the simulation at this point, which usually takes $\sim 5-10$ $t_{\rm mix}$. All temperature PDFs we show in this work are time-independent, steady-state PDFs.

\subsection{Plane-Parallel Turbulent Mixing Layer Simulation Setup} \label{sec:mixing layer sim setup}

Our turbulent mixing layer simulation setup is similar to \cite{ji19} and \cite{tan21}. The initial condition consists of half of the box at the cold stable equilibrium temperature ($T \sim 60$K for the ISM net cooling function and $T \sim 10^4$K for the CGM net cooling function) and the other half at the hot stable equilibrium temperature ($T \sim 6000$K for the ISM net cooling function and $T \sim 10^6$K for the CGM net cooling function). We introduce an initial shear velocity profile that allows the two phases to develop relative motion, which ultimately generates turbulence. To simplify the description of our setup, we define a coordinate system as follows: the x-axis is normal to the initial cold-hot interface, the y-axis is the direction of the shear flow, and the z-axis is the remaining spatial dimension. With this coordinate system, the shear velocity profile we prescribe is given by

\begin{align}
    v_y = \frac{v_{\rm shear}}{2} {\rm tanh} \left( \frac{x}{a}\right).
    \label{eq:shear velocity profile}
\end{align} 
This profile is given the following perturbation:

\begin{align}
    \delta v_x = 0.01 v_{\rm shear} {\rm exp}\left( - \frac{x^2}{a^2} \right) {\rm sin}\left( k_y z\right) {\rm sin}\left( k_z z\right).
    \label{eq:shear velocity perturbation}
\end{align}
Here, $v_{\rm shear}$ is the magnitude of the shear velocity, a is the scale length of the shear velocity profile and is set to $\sim 1\%$ of the box size. The perturbation amplitude is set to be $\sim 1\%$ of the magnitude of the shear velocity. The perturbation wavenumbers $k_y = 2\pi / \lambda_y$ and $k_z = 2\pi / \lambda_z$ are set such that the perturbation wavelengths $\lambda_y$ and $\lambda_z$ are of order the box size. The velocity perturbation $\delta v_x$ induces Kelvin-Helmholtz instability, which generates turbulence at the interface of the two phases, leading to the formation of turbulent mixing layers. Adjusting the magnitude of $v_{\rm shear}$ allows us to control the magnitude of the turbulent velocity $v_{\rm turb}$ in the mixing layer.

Additionally, we implement the relevant net cooling functions and ensure that the \Da number is greater than 1 such that radiative cooling dominates over turbulent mixing to convert hot gas into cold gas. In computing Da, we directly measure the turbulent velocity $u^{\prime}$, after subtracting out the mean shear motion. This means in the lab frame, the mixing layer is constantly consuming hot gas and shifting towards the x-boundary at the hot gas end of the box. In light of this, we adopt a frame-tracking scheme that constantly shifts to the rest frame of the mixing layer. This allows us to contain the simulation domain in a reasonably sized box and thus reduces computational cost. Frame-tracking is implemented as user-defined x-boundary conditions. All other boundary conditions in the y- and z-directions are periodic.

As with the turbulent box simulations, we run the mixing layer simulations for $\sim 5-10$ $t_{\rm mix}$ until the temperature PDF stabilizes.

\subsection{Wind Tunnel Simulation Setup}

Our wind tunnel simulation is similar to the classic setup well-known in the literature \citep{gronke18, sparre19, li20, kanjilal21, abruzzo22, farber22}. We initialize a stationary spherical cloud at $T \sim 10^4 {\rm K}$ placed in a rectangular box filled with a hot wind with initial Mach number $\mathcal{M}_{\rm wind} = \left.  v_{\rm wind}\right/ c_{\rm s,hot} = 1.5$. The cloud and the wind are in pressure equilibrium. To avoid numerical artifacts, we randomly seed percent-level density fluctuations in the initial cloud. The boundary conditions of the simulation box are all outflowing, except that we impose a constant inflow of hot gas in the wind direction. We adopt a cloud-tracking scheme that continuously shifts to the center-of-momentum frame of the cloud to reduce computational cost and prevent cloud material from leaving the simulation box. 

We run two such wind tunnel simulations with $T_{\rm hot}=10^6$K and $10^7$K, to mimic conditions in the CGM and ICM respectively. We will discuss how the different choices of $T_{\rm hot}$ affect the temperature PDF in the wind tunnel simulation in \autoref{sec:turbulent box and mixing layer}.

\begin{table*}
\caption{Summary of the Simulations We Run}
\centering
\begin{threeparttable}
\begin{tabular}{c c c c c c c c} 
\hline\hline 

\vtop{\hbox{\strut Simulation Type \tnote{a}}} &
\vtop{\hbox{\strut $T_{\rm cold}$ [K]}} & 
\vtop{\hbox{\strut $T_{\rm hot}$ [K]}} & \vline &
\vtop{\hbox{\strut \Da number=$\left. t_{\rm mix} \right/t_{\rm cool}(T_{\rm mix})$}} &
\vtop{\hbox{\strut $\mathcal{M}_{\rm turb} = \left. v_{\rm turb} \right/ c_{\rm s,hot}$}} &
\vtop{\hbox{\strut Conductivity $\kappa$ \tnote{b}}}\\ 
\hline 
ISM Turbulent Box & 60 & 6000 & \vline & 28 & 0.06 & $\kappa \propto T^{-1}$ \\
ISM Turbulent Box & 60 & 6000 & \vline & 14 & 0.13 & $\kappa \propto T^{-1}$ \\
ISM Turbulent Box & 60 & 6000 & \vline & 8 & 0.21 & $\kappa \propto T^{-1}$ \\
ISM Turbulent Box & 60 & 6000 & \vline & 5 & 0.34 & $\kappa \propto T^{-1}$ \\
ISM Turbulent Box & 60 & 6000 & \vline & 5 & 0.89 & $\kappa \propto T^{-1}$ \\
ISM Turbulent Box & 60 & 6000 & \vline & 3 & 0.55 & $\kappa \propto T^{-1}$ \\
ISM Turbulent Box & 60 & 6000 & \vline & 2 & 0.89 & $\kappa \propto T^{-1}$ \\
ISM Turbulent Box & 60 & 6000 & \vline & 28 & 0.06 & $\kappa \propto T^{-1}$, normalization boosted \\
ISM Turbulent Box & 60 & 6000 & \vline & 28 & 0.06 & $\kappa = \kappa_{\rm ISM} \propto T^{0.5}$ \tnote{c} \\
ISM Turbulent Box & 60 & 6000 & \vline & 28 & 0.06 & $\kappa = 10\kappa_{\rm ISM} \propto T^{0.5}$ \\
\hline
CGM Turbulent Box & $10^4$ & $10^6$ & \vline & 8 & 0.21 & $\kappa \propto T^{-1}$ \\
CGM Turbulent Box & $10^4$ & $10^6$ & \vline & 4 & 0.45 & $\kappa \propto T^{-1}$ \\
CGM Turbulent Box & $10^4$ & $10^6$ & \vline & 2 & 0.89 & $\kappa \propto T^{-1}$ \\
\hline
ISM Mixing Layer & 60 & 6000 & \vline & 14 & 0.13 & $\kappa \propto T^{-1}$ \\
ISM Mixing Layer & 60 & 6000 & \vline & 5 & 0.34 & $\kappa \propto T^{-1}$ \\
ISM Mixing Layer & 60 & 6000 & \vline & 2 & 0.89 & $\kappa \propto T^{-1}$ \\
ISM Mixing Layer & 60 & 6000 & \vline & 2 & 0.89 & $\kappa = \kappa_{\rm ISM} \propto T^{0.5}$ \\
ISM Mixing Layer & 60 & 6000 & \vline & 2 & 0.89 & $\kappa = 10\kappa_{\rm ISM} \propto T^{0.5}$ \\
\hline
CGM Mixing Layer & $10^4$ & $10^6$ & \vline & 8 & 0.21 & $\kappa \propto T^{-1}$ \\
\hline
ICM Mixing Layer & $10^4$ & $10^7$ & \vline & 8 & 0.21 & $\kappa \propto T^{-1}$ \\
\hline
CGM Wind Tunnel & $10^4$ & $10^6$ & \vline & N/A & N/A & $\kappa \propto T^{-1}$ \\
\hline
ICM Wind Tunnel & $10^4$ & $10^7$ & \vline & N/A & N/A & $\kappa \propto T^{-1}$ \\

\hline\hline
\end{tabular}

\begin{tablenotes}
\item[a] All simulations implement radiative cooling as we discussed in \autoref{sec:radiative cooling}.
\item[b] Our fiducial choice of the conductivity $\kappa$ is numerical diffusion (no explicit thermal conduction), which means $\kappa \propto T^{-1}$.
\item[c] $\kappa_{\rm ISM} = 2.5 \times 10^3 T^{0.5} {\rm cm}^{-1} {\rm K}^{-1} {\rm s}^{-1}$, as we discussed in \autoref{eq:kappa ISM}.
\end{tablenotes}

\label{tab:summary of simulations}
\end{threeparttable}
\end{table*}

\section{Results} \label{sec:results}


Here, we present our main simulation results. We discuss their physical interpretation in \autoref{sec:interpretation}.

\subsection{Temperature PDFs in Turbulent Box and Mixing Layer Simulations are Different} \label{sec:turbulent box and mixing layer}

\begin{figure}
\centering
\includegraphics[width=\columnwidth, height=0.8\textheight]{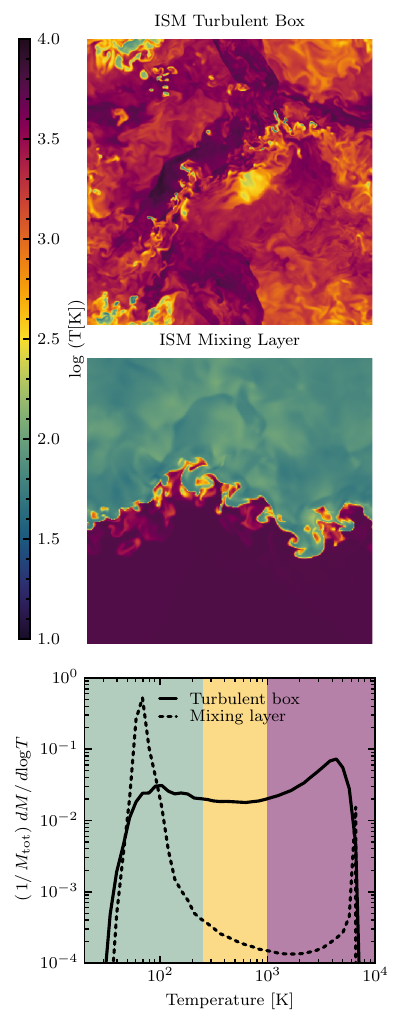}
\caption{Comparison between mixing layer and turbulent box simulations under ISM conditions. The two simulation setups here adopt identical parameters (Da=$\left. t_{\rm mix} \right/t_{\rm cool}(T_{\rm mix})=2$, $\mathcal{M}_{\rm turb}=0.89$). We compare temperature slices in the top two panels and the temperature PDFs in the bottom panel. Note that the amount of thermally stable gas (i.e. the height of the two delta functions) in the mixing layer PDF is sensitive to the box height. We discuss this effect in \autoref{sec:mixing_layer_PDF_vs_box_height} and show that it does not affect the shape of the mixing layer PDF at intermediate temperatures, which is completely different from its turbulent box counterpart.}
\label{fig:ISM_mixing_layer_vs_turb_box}
\end{figure}

In \autoref{sec:methods}, we described two simulation setups: turbulent boxes (\autoref{sec:turbulent box sim setup}) and mixing layers (\autoref{sec:mixing layer sim setup}). Naively, these two setups seem identical: they both model a multiphase, turbulent medium and use the same net cooling functions. In the turbulent mixing layer literature, a central assumption has been that zooming into a shearing interface between thermally stable phases is locally equivalent to a plane-parallel mixing layer setup, provided the thickness of the interface is small comparable to its radius of curvature. This assumption has driven considerable work on the physics of turbulent radiative mixing layers \citep{begelman90,kwak10,tan21,zhao23,sharma25}. It has been assumed that the derived temperature PDFs are universal, and can be used to derive emission and absorption line ratios \citep{tan21-lines,chen23, chenpeng26}.     

The simulation results suggest otherwise. In \autoref{fig:ISM_mixing_layer_vs_turb_box}, we show temperature slices from our turbulent box (top panel) and mixing layer (middle panel) simulations using ISM parameters and the ISM net cooling function. The mass-weighted temperature PDFs from these simulations are compared in the bottom panel. In both simulations, we fix the \Da number at ${\rm Da} = \left. t_{\rm mix} \right/ t_{\rm cool(T_{\rm mix})} = 2$ and the turbulent Mach number at $\mathcal{M}_{\rm turb}=0.89$ to eliminate confounding variables. The bottom panel of \autoref{fig:ISM_mixing_layer_vs_turb_box} shows that the temperature PDFs for turbulent box and mixing layer look completely different. When making this comparison, we are mainly interested in the shape of the PDFs at intermediate temperatures, not the amount of thermally stable gas at $\sim 60$K and $\sim 6000$K, which is arbitrarily set by the height of the simulation box in the mixing layer setup. In \autoref{sec:mixing_layer_PDF_vs_box_height}, we show that box height affects the normalization, but not the shape, of the mixing layer PDF at intermediate temperatures. The bottom panel of \autoref{fig:ISM_mixing_layer_vs_turb_box} shows that the mixing layer and turbulent box PDFs differ strongly in shape at these temperatures. The mixing layer PDF hosts most of its intermediate temperature gas mass near the cold phase at $\sim 100$K. Moving to higher intermediate temperatures, the PDF rapidly declines and levels off beyond $\sim 500$K. On the other hand, the turbulent box PDF remains flat across the entire intermediate temperature range. 

This result is surprising -- how can two simulation setups that are intended to capture the exact same physics of multi-phase turbulence with radiative cooling yield such distinct temperature PDFs? This calls into question the use of planar mixing layer simulations to make line ratio predictions in realistic turbulent flows.

\begin{figure*}
\centering
\includegraphics[width=\textwidth]{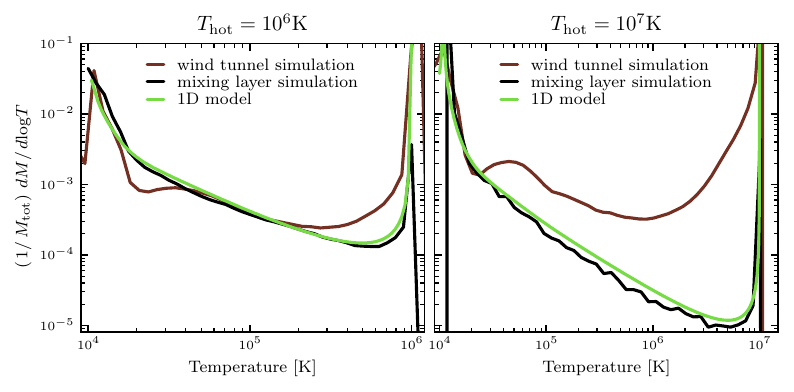}
\caption{Comparison between PDFs from wind tunnel and mixing layer simulations (normalizations of the PDFs are adjusted for easier comparison). ``1D model" refers to analytic models \citep{tan21-lines,chen23} which produce temperature PDFs that match mixing layer simulations, and are used to make line ratio predictions. Under CGM-like conditions with $T_{\rm hot} = 10^6$K, the wind tunnel and mixing layer PDFs are similar because the cold phase in the wind tunnel form an elongated cometary tail, and the intermediate temperature gas exist in concentric, roughly cylindrical shells that are well-described by mixing layer simulations and models. However, under ICM-like conditions with $T_{\rm hot} = 10^7$K, which is appropriate for jellyfish galaxy tails, the wind tunnel and mixing layer PDFs do not match. The wind tunnel PDF is less bimodal and contains more intermediate temperature gas near the hot phase, resembling the turbulent box PDF instead. This is because the density contrast in this case exceeds the critical value for shattering \citep{mccourt18, gronke20-mist}, and the cold phase exists as a collection of shattered clumps, which is not well-described by a mixing layer. These results suggest that mixing layer models for temperature PDF is by no means universal, and one needs to be careful when using mixing layer models to make line ratio predictions.}
\label{fig:temperature_PDF_cloud_vs_mixing_layer}
\end{figure*}

One might reasonably argue that while understanding the temperature PDF in the presence of volume-filling, driven extrinsic turbulence is clearly important, shear mixing layers - where turbulence is driven by Kelvin-Helmholtz and Rayleigh-Taylor instabilities -- are still ubiquitous. Quintessential examples of planar radiative turbulent mixing layers are the cometary tails of cool clouds engulfed by galactic winds, and the tails of jellyfish galaxies\citep{Yagi2010,poggianti17}. One would think these scenarios map very cleanly onto  mixing layer simulations, and produce the same temperature PDFs. To confirm this, we directly compare PDFs from wind tunnel simulations and mixing layer simulations in \autoref{fig:temperature_PDF_cloud_vs_mixing_layer}, adopting two hot phase temperatures $T_{\rm hot}=10^6$K and $10^7$K. Analytic models like \cite{tan21-lines} and \cite{chen23} yield temperature PDF predictions in green, which matches the mixing layer PDF well regardless of $T_{\rm hot}$, but the comparison between wind tunnel and mixing layer PDFs is more complicated. In the left panel, we set $T_{\rm hot}=10^6$K, a conventional choice for CGM-like conditions. Indeed, the wind tunnel PDF and mixing layer PDF are very similar. This is because in the wind tunnel simulation with $T_{\rm hot}=10^6$K, the cold cloud develops an elongated cometary tail \citep{gronke18}, and the intermediate temperature gas exist in concentric, roughly cylindrical shells that delineate this tail. This structure is well-described by a mixing layer setup. The slight difference between a cylindrical cloud surface and a plane-parallel mixing layer is reflected by the fact that the wind tunnel PDF has slightly more gas near the hot phase. However, if we increase $T_{\rm hot}$ to $10^7$K (right panel of \autoref{fig:temperature_PDF_cloud_vs_mixing_layer}), which mimics ICM conditions appropriate for jellyfish galaxy tails, things look completely different. The wind tunnel PDF with $T_{\rm hot} = 10^7$K hosts much more intermediate temperature gas near the hot phase compared to its mixing layer counterpart, with roughly equal mass per logarithmic interval. These features resemble turbulent box PDFs. The crucial difference between the two panels of \autoref{fig:temperature_PDF_cloud_vs_mixing_layer} is that with $T_{\rm hot} = 10^7$K, the density contrast between the hot and cold phases is $\left. \rho_{\rm cold} \right/ \rho_{\rm hot} = 1000$, which exceeds the critical density contrast of $\sim 300$ for shattering into a mist of tiny cloudlets \citep{mccourt18,gronke20-mist}. In a $T_{\rm hot}=10^7$K wind tunnel, the cold phase exists as a collection of shattered cold clumps instead of an elongated cometary tail, which means the conditions in the cloud tail is better described by a turbulent box rather than a mixing layer,  leading to a completely different PDF shape. The corresponding clump-in-cocoon morphology and its H$\alpha$--X-ray consequences are explored in a companion study \citep{chen26}.


The analysis above makes it clear that although the well-studied turbulent mixing layer setup adequately describe multi-phase gas dynamics in some circumstances, it is by no means "universal". A deeper understanding of the physics that sets temperature PDFs is required for charting the realm of applicability of mixing layers and understanding the regime where mixing layers might fail us.

\subsection{How Does Turbulence Strength Affect the Temperature PDF?}
\label{sec:turbulence strength and ISM PDF}

Even if we limit ourselves to just the turbulent box setup, determining the shape of the temperature PDF is still non-trivial. An important piece of physics that can affect the temperature PDF is the strength of turbulence, which varies spatially and temporally in realistic astrophysical environments. Similar to other works on ISM turbulence, we therefore vary the strength of turbulence for a box with fixed pressure and hence cooling time. However, note that this simultaneously varies two important dimensionless parameters, ${\rm Da} \propto v_{\rm turb}^{-1}$ (which dictates the relative importance of turbulent mixing and cooling), and the Mach number $\mathcal{M} \propto v_{\rm turb}$ (which dictates compressibility). Note that turbulent heating from dissipation is negligible in all our simulations.  
The sequence shown in this section should therefore be interpreted as a physically motivated progression toward stronger turbulence, rather than as an experiment isolating the effect of one parameter alone. In \autoref{sec:vary-parameters}, we briefly describe the effect of varying Da and $\mathcal{M}$ independently. 

In the absence of turbulence, all gas should settle into the thermally stable phases, and the resulting temperature PDF consists largely of two delta functions, one for each thermally stable temperature, modulo the effects of thermal conduction, which produces intermediate temperature gas at the laminar interface between warm and cold gas. Turbulence mixes these thermally stable gases and further populate the intermediate temperatures. 
Thus, we expect the abundance of intermediate temperature gas and the resulting temperature PDF to depend on the strength of turbulence: along our fiducial sequence, weak driving (large Da, small $\mathcal{M}$) should produce a relatively bimodal temperature PDF, while strong driving (small Da, large $\mathcal{M}$) should lead to a broad, flat temperature PDF. 

\begin{figure}
\centering
\includegraphics[width=\columnwidth]{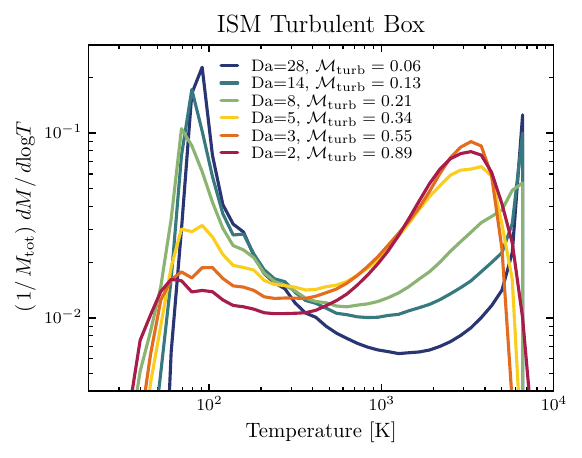}
\caption{Temperature PDFs from ISM turbulent box simulations along a fiducial turbulent-driving sequence spanning a range of \Da numbers and turbulent mach numbers. The sequence is generated by changing the turbulent velocity, so decreasing \Da number coincides with increasing turbulent mach number. Our choices of turbulent velocities ensures a sub- to trans-sonic warm ISM, which is consistent with observational constraints and existing simulations. Along this sequence, the ISM temperature PDFs evolve strongly for Da$\gtrsim5$, but saturate and adopt a fixed shape for Da$\lesssim5$. The PDFs are more bimodal at the weak-driving end of the sequence (large Da, small $\mathcal{M}$) and contain more intermediate temperature gas at the strong-driving end (small Da, large $\mathcal{M}$).}
\label{fig:ISM_turb_box_PDF_vs_Da}
\end{figure}

These ideas are verified by our ISM turbulent box simulations. \autoref{fig:ISM_turb_box_PDF_vs_Da} shows temperature PDFs of ISM turbulent box simulations at different \Da numbers and turbulent mach numbers. The \Da numbers being used here are varied by more than an order of magnitude and corresponds to turbulent velocities of 0.75, 1.5, 2.5, 4, 6.4, and 10.3 km/s in the simulation box. In comparison, the warm phase sound speed is $c_{\rm s,hot} = 11.6$ km/s. \footnote{Under our choice of $\left. P \right/ k_{\rm B} = 3000 {\rm K} {\rm cm}^{-3}$, the hot phase temperature under the ISM net cooling function is $T_{\rm hot} \sim 6000$K.} Our choices of turbulent velocities ensure that the warm ISM is subsonic to trans-sonic. This is consistent with observational constraints \citep{armstrong95, Redfield_2004, Haud2007, Gaensler2011}, which find Mach numbers of ${\mathcal M} \sim 0.5-1$ in the warm neutral medium ($T\sim 6000-10^4$K) and $\mathcal{M} \sim 1-2$ in the warm ionized medium ($T \sim 8000$K). It is also consistent with the setup of a similar suite of simulations presented in \cite{Saury:2014}. 

\autoref{fig:ISM_turb_box_PDF_vs_Da} shows that, along our fiducial turbulent-driving sequence, ISM temperature PDFs evolve strongly for ${\rm Da} \gtrsim 5$, but beyond that, the PDF saturates and adopts a fixed shape for all ${\rm Da} \lesssim 5$. We analyze the shape of these temperature PDFs from the weak-driving end of the sequence (large Da, small $\mathcal{M}$) to the strong-driving end (small Da, large $\mathcal{M}$). At large Da and small $\mathcal{M}$, cooling is expected to dominate over turbulence. The corresponding PDF (dark blue) is sharply peaked at the two thermally stable phases and skewed towards the cold phase, consistent with expectations that gas is increasingly `two-phase' as Da increases \citep{tan21}. 

As the turbulent driving is increased --- equivalently, as Da decreases and $\mathcal{M}$ increases in our fiducial suite --- the relative amount of warm to cold gas increases. The mass fraction of intermediate temperature, thermally unstable gas also increases as turbulent driving strengthens. Gas with $200 {\rm K} < T < 2000{\rm K}$ takes up $\sim 16\%$ of the total mass at the weak-driving end when Da=28. This mass fraction increases as Da decreases and reaches $\sim 36\%$ for ${\rm Da} \lesssim 5$. Beyond ${\rm Da} \lesssim 5$, the temperature PDF saturates and becomes approximately independent of further increases in driving strength along this sequence. The saturated PDF is primarily shaped by turbulence instead of cooling. It is much less bimodal, and in particular the amount of cold gas is significantly reduced. Interestingly, at low Da and high $\mathcal{M}$, the temperature PDF no longer peaks at the stable warm phase temperature $T \sim 8000$K, where net cooling vanishes, but rather at the formally unstable (in static media) temperature $T \sim 3000$K, where there is net heating. The peak near $T\sim 3000$ K at low Da and high $\mathcal{M}$ likely reflects a non-equilibrium bottleneck rather than a new stable phase: strong turbulence continuously feeds gas into the intermediate-temperature regime, while the net thermal timescale is long near $T \sim3000$ K (see \autoref{fig:net_cooling_curves_ISM_CGM}), causing gas to accumulate there. In the approximately isobaric limit $\rho \propto T^{-1}$, the mass-weighted PDF is further biased toward cooler gas. 

\begin{figure*}
\centering
\includegraphics[width=\textwidth]{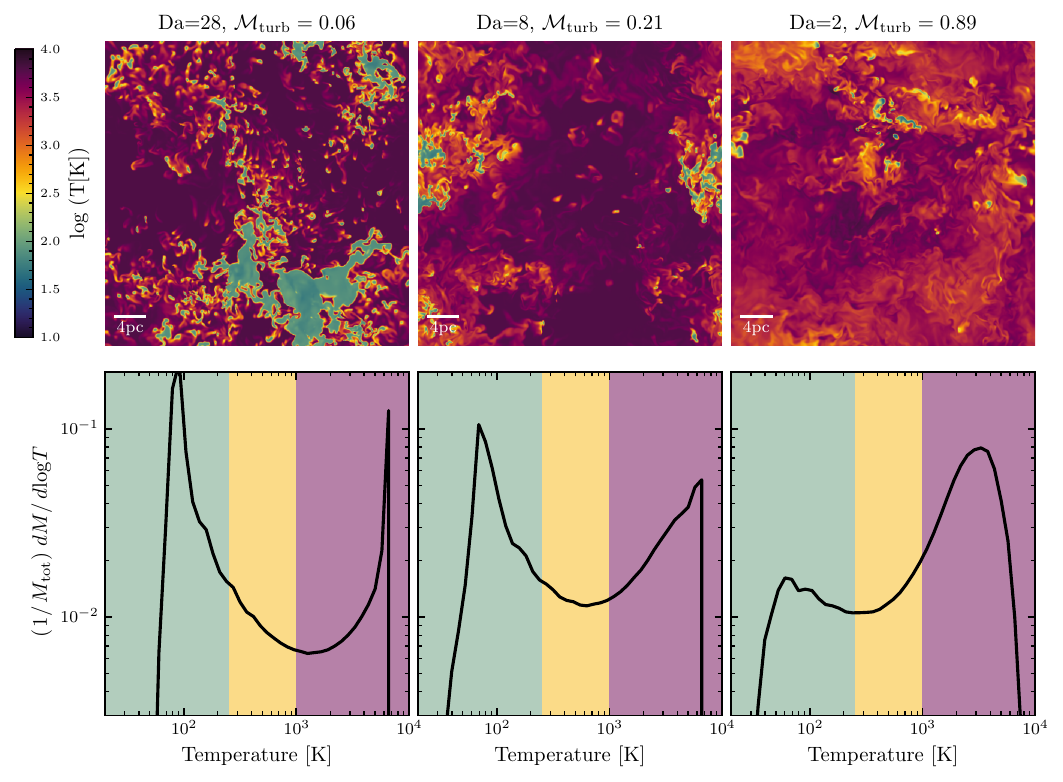}
\caption{ISM turbulent box temperature slices (\textit{top}) and temperature PDFs (\textit{bottom}) at different \Da number and turbulent mach number along the fiducial turbulent-driving sequence. Moving from left to right corresponds to increasing turbulent velocity, so decreasing \Da number coincides with increasing turbulent mach number. As turbulent velocity increases, the boundaries between the thermally stable phase (purple and blue) are blurred, and more thermally unstable gas (yellow) is generated. Stronger turbulence also breaks up cold clouds into smaller and more filamentary pieces. At Da=2 and $\mathcal{M}_{\rm turb}=0.89$, strong turbulence produces a mist of T$\sim 1000$K gas that is no longer restricted near the boundaries of the cold gas structures, and correspondingly, the temperature PDF is skewed to the hot phase.}
\label{fig:ISM_turb_box_temperature_PDF_at_different_Da}
\end{figure*}

To better understand how ISM temperature PDFs evolve along this fiducial turbulent-driving sequence, we select 3 temperature PDFs at different Da values shown in \autoref{fig:ISM_turb_box_PDF_vs_Da} and plot temperature slices of the corresponding 3D ISM turbulent box simulations in \autoref{fig:ISM_turb_box_temperature_PDF_at_different_Da}. As we move from weak to strong driving from left to right in \autoref{fig:ISM_turb_box_temperature_PDF_at_different_Da} --- equivalently, as Da decreases and $\mathcal{M}$ increases by increasing turbulent velocity --- the boundaries between the thermally stable phase (colored in purple and blue) are blurred, and more intermediate temperature, thermally unstable gas (colored in yellow) is generated. When driving is weak, intermediate temperature gas is confined near the boundary of cold clumps, but under strong driving, intermediate temperature gas can form spatially extended structures that are not limited to the vicinity of cold gas. Stronger turbulence also breaks up cold clumps into smaller and more filamentary pieces. The phenomenology of the ISM turbulent box slices is consistent with the trends we concluded from analyzing the temperature PDFs.

\begin{figure}
\centering
\includegraphics[width=\columnwidth]{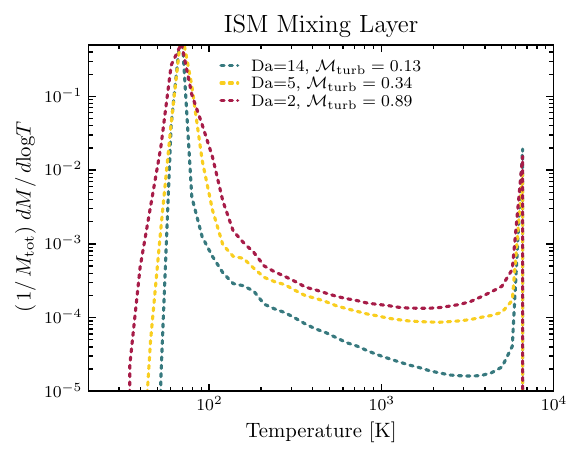}
\caption{Same as \autoref{fig:ISM_turb_box_PDF_vs_Da}, but with ISM mixing layer simulation results. Each PDF shown here is distinct from its counterpart in \autoref{fig:ISM_turb_box_PDF_vs_Da}, like we saw in \autoref{fig:ISM_mixing_layer_vs_turb_box}. The mixing layer temperature PDFs also evolve along the fiducial turbulent-driving sequence, but the dependence is weaker than what we saw with the turbulent box PDFs in \autoref{fig:ISM_turb_box_PDF_vs_Da}.}
\label{fig:ISM_mixing_layer_PDF_vs_Da}
\end{figure}

In \autoref{sec:turbulent box and mixing layer}, we saw that temperature PDFs take different shapes in turbulent boxes and mixing layers. 
How does geometry imposed by the simulation setup affect the dependence of the temperature PDFs on turbulence strength? In \autoref{fig:ISM_mixing_layer_PDF_vs_Da}, we plot ISM temperature PDFs along an analogous sequence, but for mixing layers instead of turbulent boxes. Although each individual PDF in \autoref{fig:ISM_mixing_layer_PDF_vs_Da} is different from its counterpart in \autoref{fig:ISM_turb_box_PDF_vs_Da} (which is to be expected given our findings in \autoref{sec:turbulent box and mixing layer}), the overall trend stays the same: ISM mixing layer temperature PDFs evolve with the strength of turbulence, with lower Da and higher $\mathcal{M}$ leading to more intermediate temperature gas. We note that these mixing layer PDFs are subject to the box height effect mentioned in \autoref{sec:turbulent box and mixing layer} and demonstrated in \autoref{sec:mixing_layer_PDF_vs_box_height}. However, we have kept the box height unchanged when measuring all the PDFs shown in \autoref{fig:ISM_mixing_layer_PDF_vs_Da} across mixing layers along our turbulent-driving sequence. This allows us to directly compare the abundance of intermediate temperature gas across the PDFs in \autoref{fig:ISM_mixing_layer_PDF_vs_Da}.

Several previous works reported trends of how temperature PDFs vary with the strength of turbulence. \cite{audit05} reported seeing a larger fraction of thermally unstable gas in their simulations when they increase the strength of turbulence. 
Their simulations are two-dimensional and have a different setup involving converging flows. \cite{Saury:2014} reported the same trend in their 3D ISM turbulent box simulations, but they only plotted temperature PDFs for two different turbulence strengths (see their figure 14 for details) and did not survey the parameter space comprehensively. \cite{mohapatra22-turb_box} ran turbulent box simulations under ICM conditions and also found an increase in the mass fraction of intermediate temperature gas with turbulence strength (see their figure 4 for details). These trends are generally consistent with our findings. 
However, the evolution of temperature PDFs along a turbulent-driving sequence is largely unexplained beyond the qualitative statement of "stronger turbulence implies more intermediate temperature gas". In \autoref{sec:interpretation}, we shall see that a geometric decomposition affords more insight. 

\subsection{Varying \Da and Mach numbers independently}
\label{sec:vary-parameters}

What parameters does the temperature PDF depend on? The Buckingham–$\Pi$ theorem states that if a problem depends on $n$ dimensional variables spanning $k$ independent dimensions, then its behavior can depend only on $n-k$ independent dimensionless combinations. Here, after fixing the cooling/heating law and the background pressure, the remaining macroscopic variables are $v_{\rm turb}, c_s, t_{\rm mix}$, and $t_{\rm cool}$. These span only the dimensions L and T, so $n-k=4-2=2$. The two independent dimensionless groups may be chosen as the turbulent Mach number, $\mathcal{M}=v_{\rm turb}/c_s$, and the \Da number, ${\rm Da}=t_{\rm mix}/t_{\rm cool}$. Of course, the addition of additional microphysics such as magnetic fields and conduction will introduce additional dimensionless parameters. If the pressure is allowed to vary, then it also introduces an additional dimensionless parameter (e.g., $P/P_{\rm max})$, since for a given cooling/heating law the pressure determines the underlying bistable structure.  However, in our setup, once the pressure, cooling/heating law, forcing prescription, and other microphysics are fixed, the shape of the steady normalized temperature PDF can depend only on $({\rm Da},\mathcal{M})$.

Throughout this paper, the \Da number Da = $\left. t_{\rm mix}\right/t_{\rm cool}$ is defined as an integral-scale control parameter, with $t_{\rm mix}$ evaluated on the driving (eddy turnover) scale. This is in direct analogy with the use of Da in turbulent combustion, where it is set by the integral scale of turbulence rather than by the size of individual fuel droplets, and with the survival criterion of \cite{gronke22-turb}, which is formulated in terms of the seed cloud scale rather than as an independent condition on each daughter droplet. One can in principle calculate a local effective Da for sub-box-scale structures, but as we demonstrate in \autoref{sec:scale_dependence_of_Da_and_temperature_PDF}, turbulent box temperature PDFs vary dramatically across different sub-sections of the simulation domain even on scales as large as 1/4 the box size. This means sub-box-scale Da does not map cleanly onto a fixed temperature PDF shape, which makes the integral-scale Da we use throughout this work preferable. As we will see in \autoref{sec:interpretation} and \autoref{sec:scale_dependence_of_Da_and_temperature_PDF}, the sub-box-scale variations in the temperature PDF stems from a biased sampling of key geometric features of intermediate temperature isosurfaces.

We have chosen a fixed pressure $P_{\rm eq}/k_{\rm B}\sim 3\times10^3\ {\rm K\,cm^{-3}}$, motivated by the fact that for Solar-neighborhood-like ISM conditions, a stable two-phase medium can only exist in a narrow range of pressures $P_{\rm min} < P < P_{\rm max}$, where $P_{\min}/k_{\rm B}\approx 2\times10^3\ {\rm K\,cm^{-3}}$ and $P_{\max}/k_{\rm B}\approx 5\times10^3\ {\rm K\,cm^{-3}}$ \citep{Wolfire1995}. A fixed pressure effectively fixes $t_{\rm cool}$ for a given metallicity. However, as previously noted, changing $v_{\rm turb}$ changes two dimensionless parameters at the same time: the \Da number and the turbulent mach number $\mathcal{M}_{\rm turb} = \left. v_{\rm turb} \right/ c_{\rm s,hot}$. Do they individually affect the temperature PDF, as expected from the argument above? 

\begin{figure}
\centering
\includegraphics[width=\columnwidth]{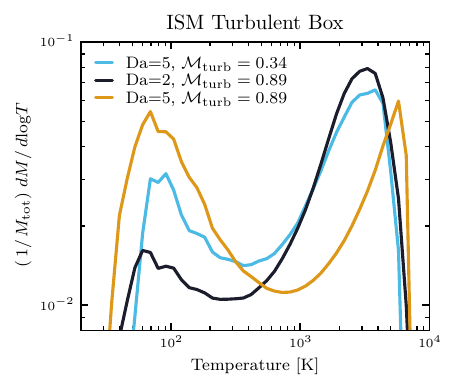}
\caption{The ISM turbulent box temperature PDFs shown here demonstrates that the shape of the temperature PDF depends on both the \Da number (Da=$\left. t_{\rm mix} \right/ t_{\rm cool}$) and  the turbulent mach number ($\mathcal{M}_{\rm turb} = \left. v_{\rm turb} \right/ c_{\rm s,hot}$). This means both of these parameters affect summary statistics like mass fraction in different ISM phases, which can be directly linked to observations. These ideas will be explored in detail in a forthcoming work.}
\label{fig:ISM_turb_box_PDF_change_Da_and_Mach_turb}
\end{figure}

We can fix the \Da number but change $\mathcal{M}_{\rm turb}$ by changing the stirring scale $L$ and the turbulent velocity $v_{\rm turb}$ by the same factor such that the eddy turnover time $t_{\rm mix} = \left. L\right/v_{\rm turb}$ and the \Da number remain unchanged. Alternatively, we can also fix $\mathcal{M}_{\rm turb}$ but change the \Da number by only changing the stirring scale $L$. These exercises should allow us to understand the roles of Da and $\mathcal{M}_{\rm turb}$ on the temperature PDF, one at a time. We demonstrate results of such experiments in \autoref{fig:ISM_turb_box_PDF_change_Da_and_Mach_turb} using ISM turbulent box simulations. In \autoref{fig:ISM_turb_box_PDF_change_Da_and_Mach_turb} the PDFs colored in blue and yellow have the same Da but different $\mathcal{M}_{\rm turb}$, while the PDFs colored in black and yellow have the same $\mathcal{M}_{\rm turb}$ but different Da. All three PDFs are distinct, indicating that indeed \textit{both} Da and $\mathcal{M}_{\rm turb}$ separately affect the shape of the temperature PDF. This is expected because $\mathcal{M}_{\rm turb}$ controls the velocity amplitude, compressibility, and turbulent deformation of cold-hot interfaces, whereas Da controls the relative timescale of cooling and mixing. Relatedly, changing the driving scale can also affect the PDF, not only by changing the effective local eddy time on cloudlet scales, but also by changing how coherently the velocity field stretches, folds, and connects the interface network. \autoref{sec:scale_dependence_of_Da_and_temperature_PDF} shows that sub-box PDFs in turbulent boxes are not universal on scales well below the driving scale, because such regions do not fairly sample the global interface geometry.


We defer a detailed exploration of higher dimensional parameter space to upcoming work. Before turning to interpretation, we consider two further empirical extensions: the role of thermal conduction and the applicability of the same picture under CGM conditions.

\subsection{Effect of Thermal Conduction}
\label{sec:conduction-results}

\begin{figure}
\centering
\includegraphics[width=\columnwidth]{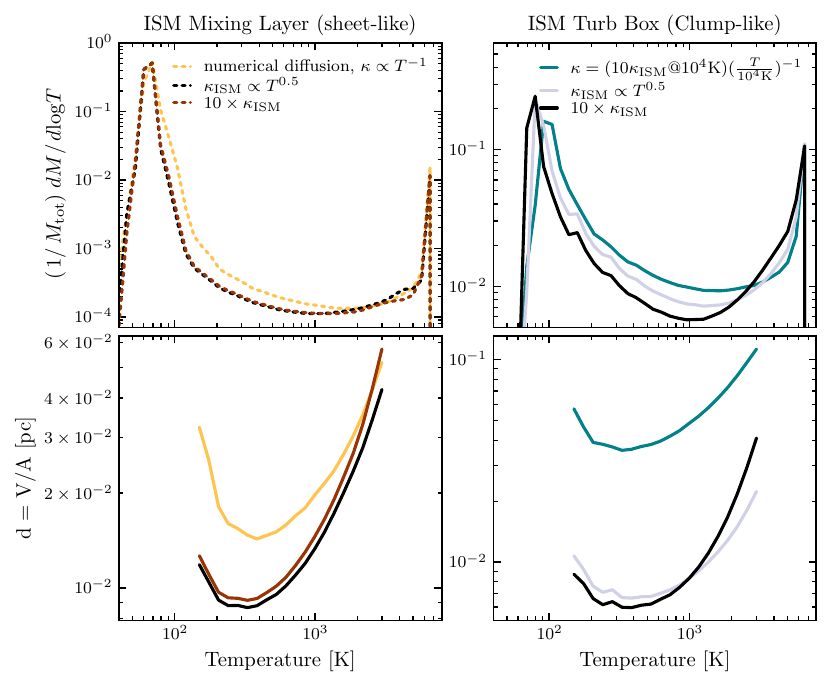}
\caption{Effect of conductivity $\kappa=aT^b$ on ISM mixing-layer (left) and weakly driven ISM turbulent-box (right) temperature PDFs. Mixing layers respond to changes in $b$ but are nearly invariant to changes in $a$. Clump-dominated turbulent boxes respond to both. The lower panels show the corresponding layer-thickness $d(T)$, which is part of the geometric  decomposition discussed in \autoref{sec:PDF-decomposition}. We further discuss how $d(T)$ is related to the temperature PDF sensitivity to the conductivity $\kappa$ in \autoref{sec:conduction-geometry}.}
\label{fig:ISM_turb_box_mixing_layer_PDF_decomposition_change_conduction}
\end{figure}

Thermal conduction provides a useful test of how much of the temperature PDF is controlled by microphysics versus morphology. We write the conductivity as $\kappa=aT^b$, where $a$ sets the normalization and $b$ the temperature dependence. Previous, in CGM mixing layers, \citet{tan21-lines} found that the PDF changes when $b$ is changed, but is nearly unchanged when only $a$ is varied. Our ISM mixing layers reproduce this result\footnote{The visual change in \autoref{fig:ISM_turb_box_mixing_layer_PDF_decomposition_change_conduction} is milder than in \citet{tan21-lines}. Besides the change in cooling function, we also plot mass-weighted PDFs rather than volume-weighted ones, which further compresses hot-side differences in an approximately isobaric flow, accounting for this difference.}(\autoref{fig:ISM_turb_box_mixing_layer_PDF_decomposition_change_conduction}, left). \citet{tan21-lines} interpreted their result in terms of a 1D analytic model for a planar conduction front, which is sensitive to the temperature dependence but not the normalization of conduction. This model could reproduce simulation results extremely well. 

The turbulent-box result is different. In weakly driven ISM boxes, where the intermediate-temperature gas is confined to the boundaries around cold clumps, changing either $a$ or $b$ changes the PDF (\autoref{fig:ISM_turb_box_mixing_layer_PDF_decomposition_change_conduction}, right). In particular, boosting the normalization of $\kappa$ shifts mass toward hotter intermediate temperatures and broadens the PDF, even when the temperature dependence is held fixed. Conductivity normalization is therefore not universally irrelevant; its importance depends on the interface morphology.

\begin{figure}
\centering
\includegraphics[width=\columnwidth]{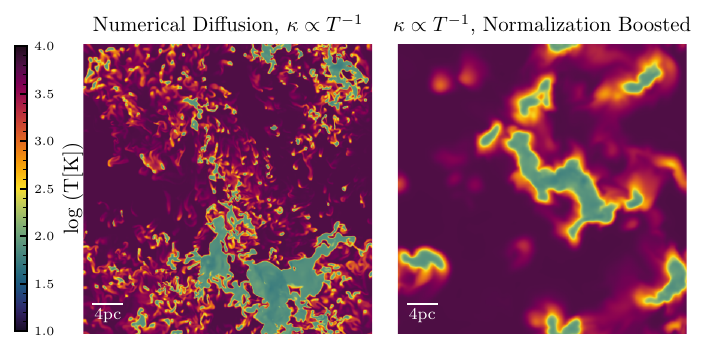}
\caption{Temperature slices from two ISM turbulent-box runs that differ only in the normalization of $\kappa$. Larger conductivity smooths the cold phase and erases the smallest structures, increasing the minimum cloud size from sub-pc scales to $\sim 5$ pc.}
\label{fig:ISM_turb_box_temperature_slice_with_different_conduction}
\end{figure}

A larger conductivity also smooths the cold phase on small scales. \autoref{fig:ISM_turb_box_temperature_slice_with_different_conduction} shows that increasing $a$ erases the smallest cold structures and increases the minimum cloud size from sub-pc scales to $\sim 5$ pc. The effect of conduction therefore cannot be reduced to a universal mixing-layer rule; we explain the geometric origin of this difference in \autoref{sec:conduction-geometry}.

\subsection{Extension to CGM Conditions}
\label{sec:CGM-extension}

\begin{figure}
\centering
\includegraphics[width=\columnwidth]{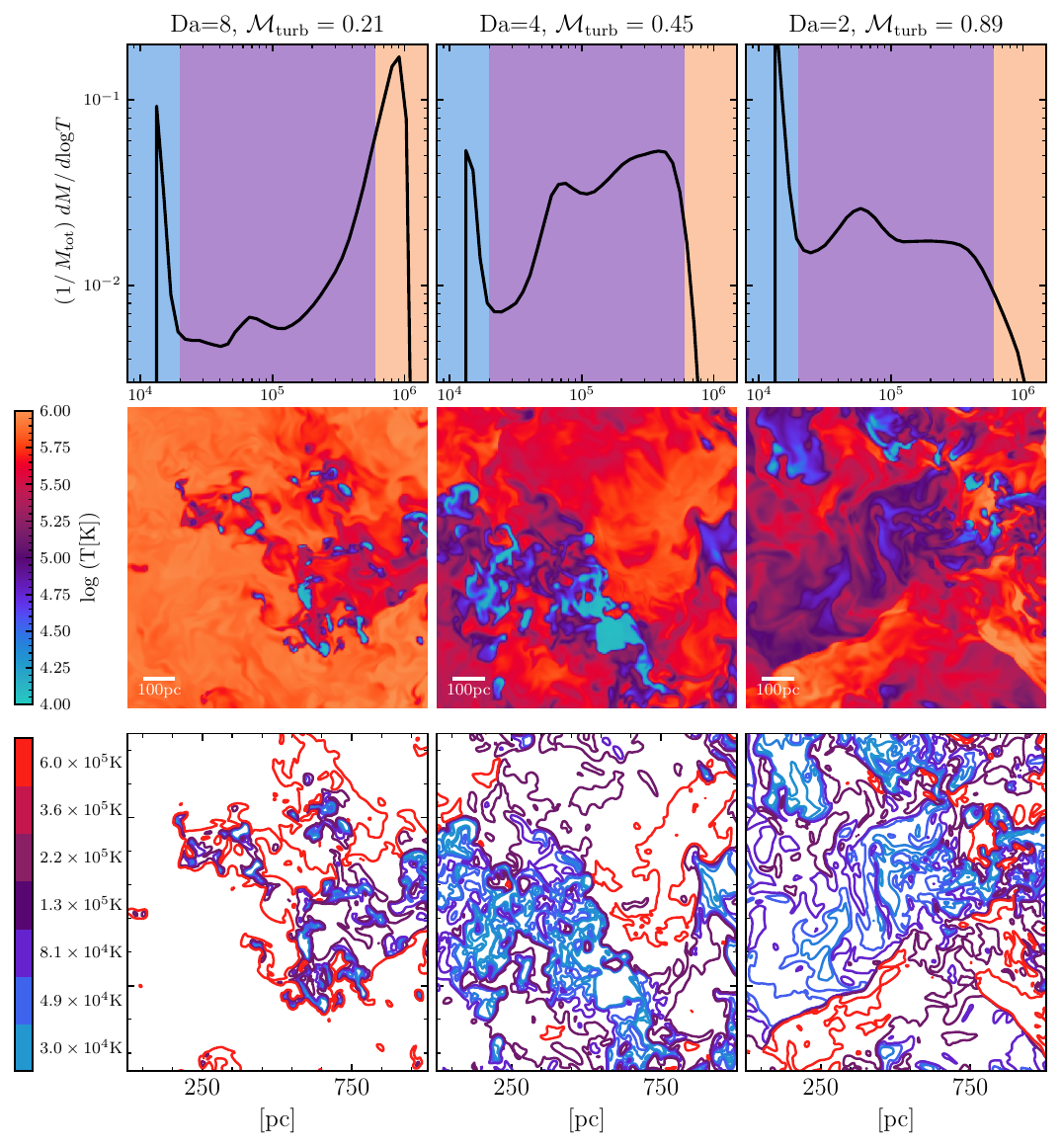}
\caption{CGM turbulent-box runs along a driving sequence analogous to the ISM suite. Weak driving gives clearly defined boundaries between the hot and cold phases and relatively bimodal PDFs; strong driving produces spatially extended intermediate temperature structures and broader PDFs. These trends are identical to ISM turbulent box results presented in \autoref{fig:ISM_turb_box_temperature_PDF_at_different_Da} and will be explained in \autoref{sec:interpretation} through a geometric decomposition of the temperature PDFs. Bottom row shows temperature contours similar to \autoref{fig:ISM_turb_box_temperature_contours} and is useful for visualizing morphology of intermediate temperature isosurfaces (see \autoref{sec:morphology} for details).}
\label{fig:CGM_turb_box_temperature_PDF_at_different_Da}
\end{figure}

The ISM suite establishes the basic picture, but the same issues arise in the CGM. \autoref{fig:CGM_turb_box_temperature_PDF_at_different_Da} shows CGM turbulent boxes along a driving sequence analogous to the ISM runs. The PDFs again evolve from relatively bimodal at weak driving to broader at strong driving. The temperature slices show that as driving strengthens, the boundaries between thermally stable phases are blurred, and intermediate temperature gas can start to form spatially extended structures instead of being confined near the cold phase. These behaviors are identical to the ISM turbulent box results presented in \autoref{sec:turbulence strength and ISM PDF}. How can we systematically understand these trends?


\section{Understanding Temperature PDFs through Morphology}
\label{sec:interpretation}

\subsection{Surface Area and Thickness Decomposition of Temperature PDFs}
\label{sec:PDF-decomposition}


How should one interpret the temperature PDF? At the most fundamental level, the volume of gas occupying a narrow temperature interval $[T,T+dT]$ is a geometric quantity. Let $\Sigma_T$ denote the temperature isosurface. The co-area formula, originating in geometric measure theory, is a generalization of the change-of-variables formula that relates integrals over a volume to integrals over the isosurfaces of a scalar field \citep{evans25}. It gives:
\begin{equation}
 \frac{dV}{dT}=\int_{\Sigma_T}\frac{dS}{|\nabla T|},
\end{equation}
which is simply the statement that the volume between two nearby level surfaces $T$ and $T+dT$ equals the isosurface area times an effective normal thickness $dT/|\nabla T|$. Applied here, it converts the temperature PDF into a geometric statement about temperature isosurfaces. Motivated by this, we introduce the more intuitive finite-bin decomposition: 
\begin{align}
    V_{\rm tot,T} = A_{\rm isosurface,T} \times d_{(T)}, 
    \label{eq:area_thickness_decompoisition}
\end{align} 
where $A_{\rm isosurface,T}$ is the area of the temperature isosurface, and $d_{(T)} = \left. V_{\rm tot,T} \right/ A_{\rm isosurface,T}$ represents a characteristic thickness of that temperature layer. 
This decomposition allows us to distinguish between two physically different effects: $d(T)$, which is controlled primarily by microphysical processes such as radiative cooling, heating, and thermal conduction, and $A(T)$, which is controlled by the morphology of the multiphase structure. Since our simulations are close to isobaric, the mass-weighted temperature PDF differs from the volume-weighted one only by the usual isobaric factor $\rho\propto T^{-1}$, so the same geometric decomposition also provides a natural interpretation of the mass PDF shown throughout this paper. While  our decomposition of $V_{\rm tot,T}$ might appear as a tautology, the key is to discern whether variations in $V(T)$ are due to changes in $A(T)$ or $d(T)$; these changes have different physical origins. 

\begin{figure}
\centering
\includegraphics[width=\columnwidth]{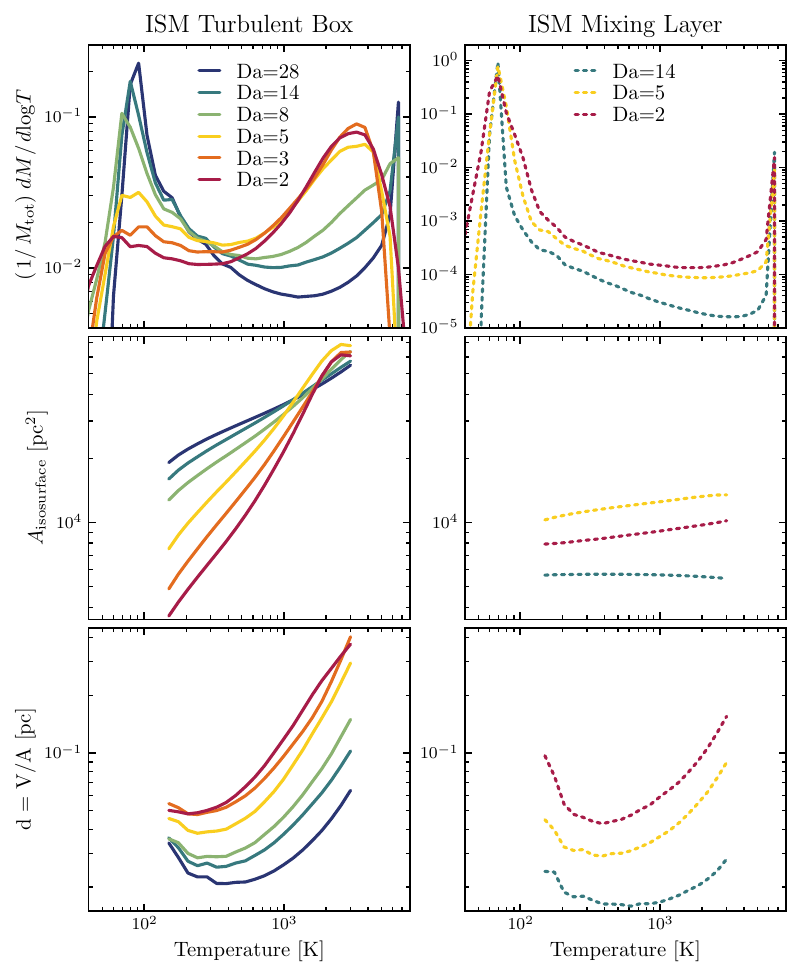}
\caption{Geometric decomposition of ISM turbulent box and mixing layer temperature PDFs (\textit{top}) into the product of temperature isosurface area (\textit{middle}) and isosurface thickness (\textit{bottom}), according to \autoref{eq:area_thickness_decompoisition}. The isosurface thickness as a function of temperature ($d_{(T)}$, \textit{bottom}) encodes micro-physical processes like cooling, heating, and conduction and does not depend strongly on turbulence strength for both turbulent box and mixing layer (up to normalization). The isosurface area as a function of temperature ($A_{(T)}$, \textit{middle}) encodes the morphology of intermediate temperature interfaces and shows a stronger dependence on turbulence strength along our fiducial turbulent-driving sequence. In particular, $A_{(T)}$ are roughly power-laws regardless of the simulation setup, but the power-law is much steeper for turbulent boxes, which explains the different temperature PDFs across turbulent boxes and mixing layers. Looking at just the turbulent box PDF decomposition, the different power-law slopes of $A_{(T)}$ are responsible for the sensitivity of the PDFs to the strength of turbulence driving. Interestingly, $A_{(T)}$ for turbulent boxes shows a flattening at high T and small Da. This hints at a transition from clump-like to sheet-like geometry, which will be explored extensively in the following sections and figures.}
\label{fig:ISM_turb_box_mixing_layer_PDF_decomposition}
\end{figure}

In \autoref{fig:ISM_turb_box_mixing_layer_PDF_decomposition}, we decompose temperature PDFs (top row) into isosurface areas and characteristic thicknesses (middle and bottom rows, both as a function of temperature) for both the ISM turbulent box and mixing layer simulations at a range of turbulent strengths, which was discussed in \autoref{sec:results}. To compute the surface area of temperature isosurfaces, we utilize the marching cubes algorithm \citep{lorensen87} implemented in the python package \texttt{scikit-image} \citep{vanderWalt14}. We demonstrate in \autoref{sec:time_evolution_of_A_d_decomposition} that this area-thickness decomposition is time-independent once the temperature PDF reaches steady-state.

\autoref{fig:ISM_turb_box_mixing_layer_PDF_decomposition} already contains the main geometric result of this paper: the interesting PDF differences are driven primarily by $A(T)$, not $d(T)$. In the turbulent-box suite (left column), the shape of $A_{(T)}$ evolves strongly from the weak-driving end of the sequence (large Da, small $M_{\rm turb}$) to the strong-driving end (small Da, large $M_{\rm turb}$), whereas the shape of $d_{(T)}$ changes much less.\footnote{In practice, the plotted $d_{(T)}$ is a finite-bin estimator of a differential thickness. While its normalization is sensitive to binning, its shape is converged with respect to both temperature binning and numerical resolution (\autoref{sec:convergence test}).} Since the PDFs are normalized, an overall puffing-up of all temperature layers changes the normalization of $d_{(T)}$ but not the PDF shape. Thus, the strong evolution of the turbulent-box PDFs along the fiducial sequence is primarily geometric: it reflects changes in $A_{(T)}$. At $T\sim 300\,{\rm K}$, the isosurface area is larger on the weak-driving end of the sequence, which produces a larger low-$T$ peak in the PDF. At $T\sim 2000\,{\rm K}$, the trend reverses: the isosurface area is larger on the strong-driving end, and correspondingly more mass resides at higher temperatures. 

The same figure also explains why turbulent-box and mixing-layer PDFs differ under the same ISM conditions. Cross-comparing the two columns of \autoref{fig:ISM_turb_box_mixing_layer_PDF_decomposition}, both $A_{(T)}$ and $d_{(T)}$ differ between the two setups, but the difference in $A_{(T)}$ is much more pronounced and dominates the PDF shape. In particular, $A_{(T)}$ depends much more strongly on temperature in turbulent boxes than in mixing layers. This is why turbulent boxes produce flatter, broader PDFs with more intermediate-temperature gas, whereas mixing layers remain much more bimodal.

By contrast, the mixing-layer PDFs evolve only weakly along the fiducial sequence because $A_{(T)}$ is almost fixed by the imposed planar geometry. In that case the modest changes in the PDF are controlled primarily by modest changes in $d_{(T)}$. This is precisely why planar mixing-layer models are so successful at capturing interface microphysics, yet fail to reproduce the full temperature distribution once the geometry of the cold gas becomes non-trivial.

\subsection{Understanding Surface Area through Morphology}
\label{sec:morphology}

\begin{figure*}
\centering
\includegraphics[width=\textwidth]{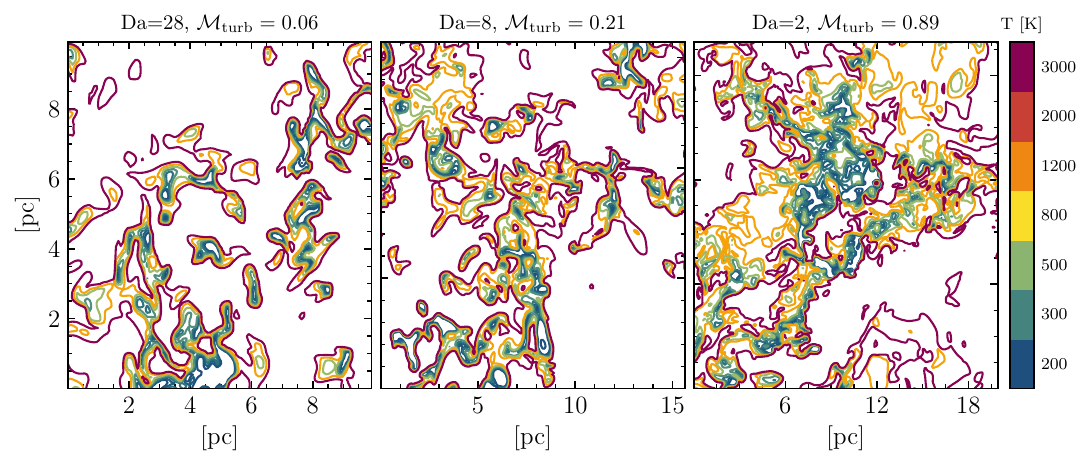}
\caption{Temperature contours in ISM turbulent boxes along the fiducial driving sequence. Each panel shows a representative region in a slice of the simulation box. At weak driving, the contours form onion-skin shells around cold clumps. At strong driving, the low-$T$ contours remain clump-like but the $T \gtrsim 800$K contours merge into extended sheets. This clump-to-sheet transition explains the change in $A_{(T)}$ seen in \autoref{fig:ISM_turb_box_mixing_layer_PDF_decomposition}.}
\label{fig:ISM_turb_box_temperature_contours}
\end{figure*}

The decomposition in \autoref{sec:PDF-decomposition} reduces the problem to a sharper one: what sets $A_{(T)}$? In turbulent boxes, the answer is morphology. Intermediate-temperature isosurfaces trace the skins of cold clouds, so their area depends on whether those skins remain isolated or merge into extended networks.

The slice plots in \autoref{fig:ISM_turb_box_temperature_PDF_at_different_Da} already show the trend. At weak driving (large Da, small $\mathcal{M}_{\rm turb}$), cold gas resides in distinct clumps and the intermediate-temperature gas forms thin shells around them. At strong driving (small Da, large $\mathcal{M}_{\rm turb}$), those shells connect through the box and produce a diffuse $T\sim 10^3$K ``mist.'' The geometry therefore shifts from clump-like to sheet-like.

This is even clearer in the contour plots of \autoref{fig:ISM_turb_box_temperature_contours}. Along the driving sequence, isolated closed contours give way to large connected structures. At fixed strong driving, morphology also depends on temperature: the lower-$T$ contours ($T \lesssim 500$K) still wrap individual clouds, whereas the higher-$T$ contours ($T \gtrsim 800$K) already percolate into sheets. At that point the largest structures reach order-unity area covering fractions.

\begin{figure*}
\centering
\includegraphics[width=\textwidth]{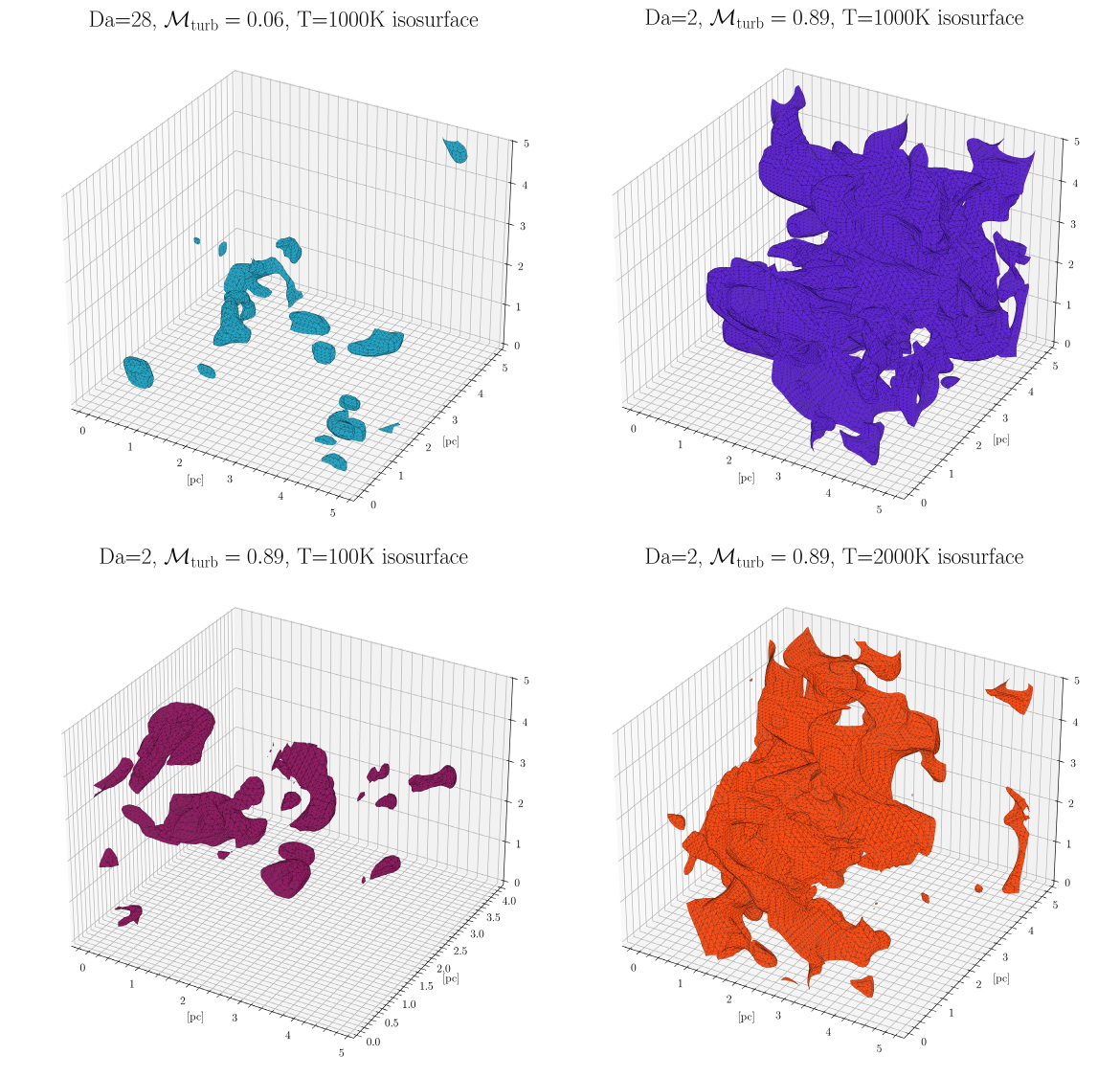}
\caption{3D temperature isosurfaces illustrating the same morphological transition. Each panel shows a representative subsection of the entire simulation box. \textit{Top}: at $T=1000$K, weak driving gives disconnected clumps, while strong driving gives an extended sheet. \textit{Bottom}: in the strong-driving run, the $100$K isosurface is clump-like but the $2000$K isosurface is sheet-like. Stronger driving and higher temperature both favor sheets.}
\label{fig:clump_vs_sheets_3d}
\end{figure*}

\autoref{fig:clump_vs_sheets_3d} shows the same transition in 3D. At $T=1000$K, the weak-driving run consists of disconnected shells, while the strong-driving run contains a single extended sheet. Within the strong-driving run, the $100$K isosurface remains clumpy but the $2000$K isosurface is sheet-like. Thus both stronger driving and higher temperature push the interface toward connected sheets.

This immediately explains the form of $A_{(T)}$ in \autoref{fig:ISM_turb_box_mixing_layer_PDF_decomposition}. In the clump-dominated regime, isosurfaces behave like onion skins: hotter shells lie at larger radii and therefore have larger area, producing the roughly power-law rise of $A_{(T)}$ with $T$. Under strong driving, the low-$T$ layers are still clumpy, but the higher-$T$ layers merge across neighboring clouds. That connectivity boosts the high-$T$ area and broadens the PDF. Once topology becomes sheet-like, then $A(T)$ becomes independent of temperature. This flattens $A_{(T)}$ at the highest temperatures.

\begin{figure}
\centering
\includegraphics[width=\columnwidth]{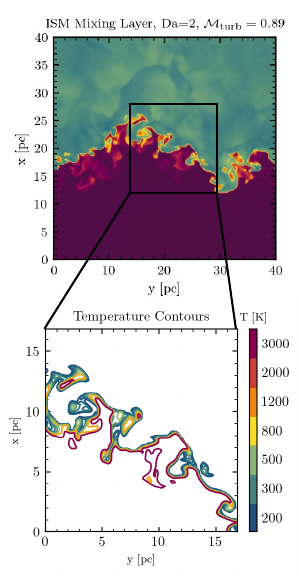}
\caption{Temperature contours in the ISM mixing-layer run with Da=2 and $\mathcal{M}_{\rm turb}=0.89$ (\textit{Bottom}). For better visualization, we only show temperature contours of a zoomed-in section of the simulation box. The full mixing layer interface is shown in the \textit{top} panel to help with contextualization. Shear motion is along the y-direction (horizontal axis of both panels), as described in \autoref{sec:mixing layer sim setup}. In this mixing layer setup, the temperature contours remain stacked sheets, so $A_{(T)}$ depends only weakly on temperature or driving. The PDF therefore varies mainly through the layer thickness $d_{(T)}$.}
\label{fig:ISM_mixing_layer_Da_2_temperature_contours}
\end{figure}

None of this occurs in mixing layers. There the geometry is imposed rather than emergent. As \autoref{fig:ISM_mixing_layer_Da_2_temperature_contours} shows, the isosurfaces remain stacked sheets, deformed by Kelvin--Helmholtz motions but never broken into clumps. Consequently, $A_{(T)}$ depends only weakly on $T$, the \Da number, or $\mathcal{M}_{\rm turb}$, and the modest PDF variations are driven mainly by $d_{(T)}$. This is why planar mixing-layer temperature PDF depends primarily on interface microphysics such as the interplay between cooling, enthalpy advection and thermal conduction, which set $d_{\rm (T)}$ \citep{tan21-lines}, yet miss the full temperature distribution once the cold phase develops non-trivial morphology.

\begin{figure}
\centering
\includegraphics[width=\columnwidth]{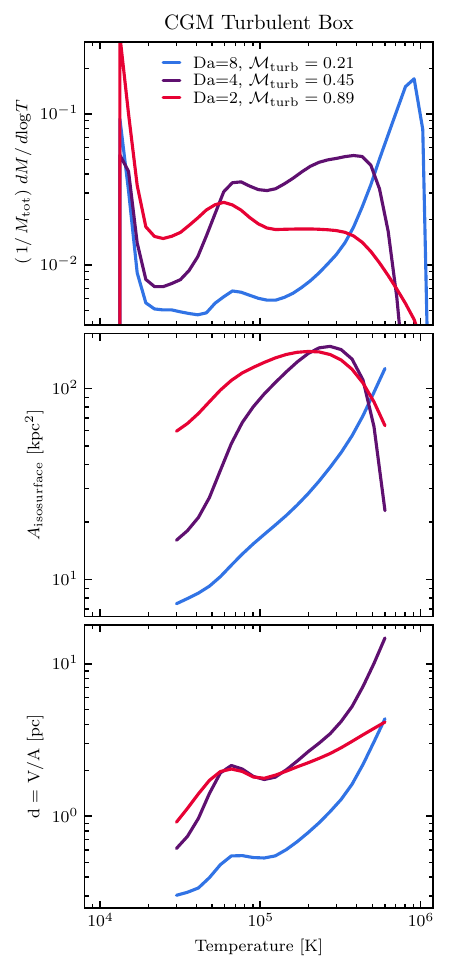}
\caption{Diagnostic decomposition of the CGM turbulent-box PDFs shown in \autoref{fig:CGM_turb_box_temperature_PDF_at_different_Da}. As in the ISM suite, most of the variation tracks the temperature dependence of the isosurface area $A_{(T)}$, while the layer thickness $d_{(T)}$ changes much less.}
\label{fig:CGM_turb_box_PDF_decomposition}
\end{figure}

The CGM results in \autoref{sec:CGM-extension} suggest that the geometric picture developed above is not peculiar to the ISM. Despite the different temperature scale and cooling curve, the same morphological progression from isolated clumps to connected sheets reappears, as shown by the temperature slices and contour plots in \autoref{fig:CGM_turb_box_temperature_PDF_at_different_Da}. Similar to \autoref{fig:ISM_turb_box_mixing_layer_PDF_decomposition}, we provide a diagnostic decomposition of the CGM PDFs in \autoref{fig:CGM_turb_box_PDF_decomposition}, which points to the same control parameter as in the ISM: most of the variation tracks the temperature dependence of the isosurface area, not the layer thickness. At weak driving, $A_{(T)}$ rises roughly as a power law, consistent with clump-like onion-skin interfaces. At strong driving, $A_{(T)}$ turns over above $T \gtrsim 2\times 10^5$K, indicating that the hotter isosurfaces are percolating into connected sheets\footnote{For the strongest driving (Da=2), the CGM curves for $d(T)=V/A$ are not strictly self-similar with the other curves. We would not over interpret this. The more robust result is that the evolution of the CGM PDF is still dominated by changes in $A(T)$, in particular its high-temperature turnover as the morphology transitions from clump-like to sheet-like.}

Thus, the clump-to-sheet transition and its imprint on the temperature PDF are not specific to the ISM. The same geometric control appears under CGM cooling and at much higher absolute temperatures. We therefore view the clump-to-sheet transition as a generic organizing principle for multiphase turbulence whenever the cold phase fragments rather than remaining in a single planar layer.

\subsection{Quantitative Description of Morphology}
\label{sec:betti numbers}

\begin{figure}
\centering
\includegraphics[width=\columnwidth]{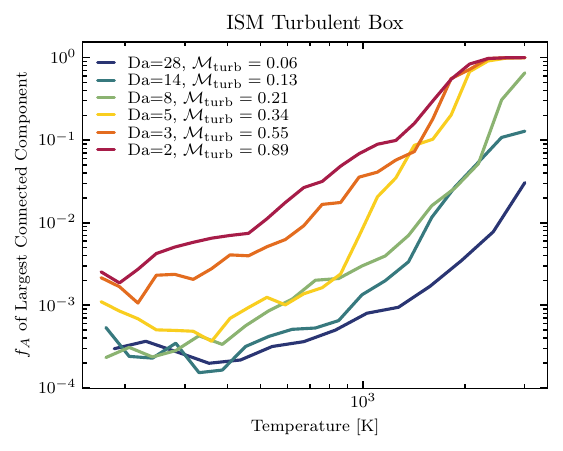}
\caption{Area covering fraction $f_A$ of the largest connected component. $f_A$ rises with temperature and driving strength, showing the transition from disconnected clumps to percolating sheets.}
\label{fig:ISM_turb_box_largest_component_f_A}
\end{figure}

The visual argument above can be made quantitative. We first measure the area covering fraction $f_A$ of the \textit{largest} connected component of each isosurface, identifying connected regions with the python package \texttt{connected-components-3d} \citep{cc3d_2021}. Clump-like morphology should have small $f_A$; sheet-like morphology should have large $f_A$. \autoref{fig:ISM_turb_box_largest_component_f_A} shows exactly this: $f_A$ increases with both temperature and driving strength, and approaches unity at high $T$ in the strongly driven runs. The same trend seen by eye is therefore present in a simple connectivity statistic.

\begin{figure}
\centering
\includegraphics[width=\columnwidth]{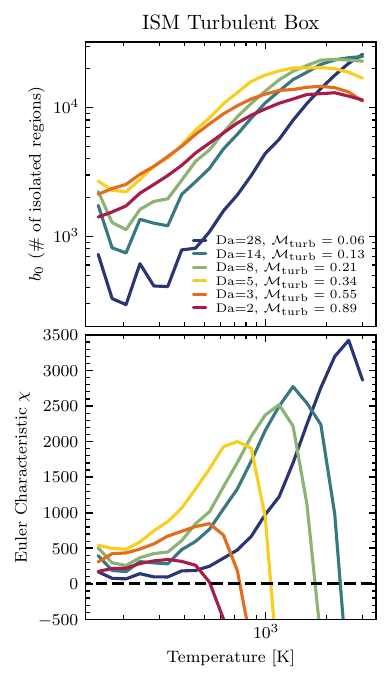}
\caption{\textit{Top}: Betti number $b_0$, the number of isolated components. Weakly driven runs show a monotonic rise with temperature, while strongly driven runs turn over at high $T$ as clumps merge into sheets. \textit{Bottom}: Euler characteristic $\chi$; $\chi>0$ indicates clump-dominated geometry and $\chi<0$ sheet-dominated geometry. The zero-crossing moves to lower temperature as the driving strengthens. Once $\chi$ drops below 0, it monotonically decreases with $T$.}
\label{fig:ISM_turb_box_b0_euler_characteristic}
\end{figure}

A more direct topological diagnostic is the Betti number $b_0$, computed with \texttt{quantimpy} \citep{quantimpy}, which is the number of isolated components. In weakly driven boxes, $b_0$ rises monotonically with temperature. In strongly driven boxes, by contrast, $b_0$ eventually plateaus and then declines at high $T$ (\autoref{fig:ISM_turb_box_b0_euler_characteristic}). In the strict nested-shell (onion skin) picture, $b_0$ should be nearly temperature-independent. Its rise with temperature indicates that warmer interface gas is distributed around a larger number of distinct structures than the coldest gas-- that is, interface regions appear that never cool down to the coldest temperatures\footnote{There is also some sensitivity to temperature bin-size: narrow temperature bins can break a geometrically continuous interface into several disconnected patches.}. At higher temperatures and in strongly driven boxes, those structures merge into a dominant connected sheet. That turnover of $b_{0(T)}$ in strongly driven boxes marks the onset of percolation: neighboring shells merge into connected sheets, so the number of isolated pieces falls even as the occupied volume continues to grow. The same merger of temperature isosurfaces also explains the mild high-$T$ flattening of $A_{(T)}$ in the strong-driving runs (middle panel of left column in \autoref{fig:ISM_turb_box_mixing_layer_PDF_decomposition}), since one connected sheet has less total area than many disconnected shells with the same combined volume.

Besides $b_0$, it is useful to consider higher order Betti numbers. While $b_0$ counts disconnected components, $b_1$ measures the number of non-contractible loops, i.e. the genus or handle complexity of the surface. In the clump-dominated regime, structures are approximately simply connected ($b_1 \approx 0$), whereas the transition to a connected, sheet-like interface proceeds primarily through the formation of handles via folding and reconnection, driving a rapid increase in $b_1$. The second Betti number $b_2$, which counts fully enclosed cavities, remains negligible throughout. This reflects the fact that the multiphase interface is predominantly open and percolating, rather than forming closed shells that isolate volumes of one phase within another. As a result, the Euler characteristic: 
\begin{align}
    \chi=b_0-b_1+b_2, 
    \label{eq:euler characteristic}
\end{align}
is well approximated by $\chi \sim b_0 - b_1$. For our purposes, $\chi>0$ indicates clump-dominated geometry, while $\chi<0$ indicates sheet-dominated geometry. The bottom panel of \autoref{fig:ISM_turb_box_b0_euler_characteristic} shows that weakly driven runs remain at $\chi>0$ throughout the intermediate-temperature range, i.e., they remain clump-like. As the driving strengthens, the zero-crossing moves to lower temperature, meaning the clump-to-sheet transition happens earlier. At the strong driving end (${\rm Da} \lesssim 5$), sheet-like geometry dominates beyond the zero crossing of $\chi$ at $T \lesssim 1000$K. This explains why the temperature PDFs along our turbulent-driving sequence shown in \autoref{fig:ISM_turb_box_PDF_vs_Da} saturates and adopts a fixed shape for ${\rm Da} \lesssim 5$ and $T \gtrsim 1000$K.

\begin{figure}
\centering
\includegraphics[width=\columnwidth]{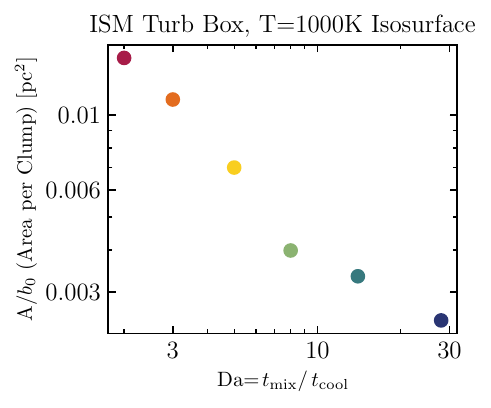}
\caption{Area per isolated component, $A/b_0$, evaluated at $T=1000$K. It rises as the \Da number decreases (as turbulence driving strengthens), consistent with fewer, larger connected structures and a shift toward sheet-like morphology.}
\label{fig:ISM_surface_area_per_clump_at_intermediate_T}
\end{figure}

A final summary statistic is $A/b_0$, the area per isolated component. At $T=1000$K, this quantity rises as turbulent driving becomes stronger (\autoref{fig:ISM_surface_area_per_clump_at_intermediate_T}), showing that the connected structures become larger and less numerous.

Taken together, $f_A$, $b_0$, $\chi$, and $A/b_0$ all support the same picture: stronger driving and higher temperature transform a set of disconnected cloud skins into percolating sheets, and that geometric transition is what reshapes the temperature PDF.

In this work, we have used the \Da number Da=$\left. t_{\rm mix} \right/ t_{\rm cool}$ to characterize the strength of turbulence driving and analyzed how temperature PDFs are sensitive to Da in \autoref{sec:results}. We consistently use the integral-scale Da by evaluating $t_{\rm mix}$ on the driving scale and justified this choice in \autoref{sec:vary-parameters}. It is worth noting that Da is scale-dependent, and the local effective Da on the scale of individual cloudlets is generally smaller than the integral-scale value and varies across the cold-phase size hierarchy, which is approximately scale-free ($\left. {\rm dN} \right/ {\rm dm} \propto {\rm m}^{-2}$; \citealt{gronke22-turb}). However, this scale-dependence is not a competing explanation for the temperature PDFs we present here. \cite{tan21-lines} found that changing quantities that modify the cooling time (and hence Da), such as the normalization of the cooling function, left the temperature PDF nearly unchanged. This is consistent with findings in our ISM mixing layer simulation suite, where $d_{(T)}$ mediates slight changes in the temperature PDF that are much less dramatic compared to variations in the turbulent box suite (\autoref{fig:ISM_turb_box_mixing_layer_PDF_decomposition}). This means any role for scale-dependent local Da must therefore be expressed through $A_{(T)}$, which the geometric framework of \autoref{sec:interpretation} measures directly. The clump-to-sheet transition may equivalently be regarded as the morphological consequence of an effective Da that decreases on smaller scales; the value of the geometric description is that it captures this transition through quantities — $A_{(T)}$, the largest-component area fraction $f_A$, the Betti number $b_0$, the Euler characteristic $\chi$, and the area per isolated component $\left. A \right/ b_0$ — that map directly onto the PDF.

\subsection{Why Conduction Behaves Differently in Clump-like and Sheet-like Geometries}
\label{sec:conduction-geometry}

The conduction results in \autoref{sec:conduction-results} are a direct test of the $A(T)d(T)$ framework. Writing $\kappa=aT^b$, changing $b$ alters the relative thickness of the temperature layers at different $T$. It therefore reshapes $d_{(T)}$ and changes the PDF in both mixing layers and turbulent boxes, as shown in the bottom row of \autoref{fig:ISM_turb_box_mixing_layer_PDF_decomposition_change_conduction}.

Changing only the normalization $a$ is different. To leading order it multiplies the thickness of every layer by nearly the same factor. For a sheet-like interface, that simply moves a stack of nearly parallel isosurfaces apart by a common scale factor: the layers become thicker, but their relative amounts hardly change, so the normalized PDF is almost invariant. This is the special case captured by planar mixing-layer models and shown in the bottom left panel of \autoref{fig:ISM_turb_box_mixing_layer_PDF_decomposition_change_conduction}.

In a clump-like morphology, however, the same puffing-up acts on concentric shells around cold clouds. Thickening those shells changes not only $d_{(T)}$ but also the shell radii. Because the hotter shells sit at larger radii, they gain more area and volume than the colder ones, so the relative occupancy of temperature bins changes and the PDF becomes sensitive to $a$. Stronger conduction also smooths away the smallest cold clumps (\autoref{fig:ISM_turb_box_temperature_slice_with_different_conduction}), modifying $A_{(T)}$ itself. The normalization of $\kappa$ therefore matters in clump-dominated flows for two geometric reasons: it inflates onion-skin shells and it erases small-scale structure. The result of \cite{tan21-lines} is therefore a sheet-geometry limit, not a universal property of conduction. In practice, this normalization dependence is visible only when the conductive skin is resolved by at least a few grid cells and is shown in the bottom right panel of \autoref{fig:ISM_turb_box_mixing_layer_PDF_decomposition_change_conduction}.

\section{Discussion and Summary}
\label{sec:summary}

The central result of this paper is that temperature PDFs are set by both microphysics and geometry. Existing mixing-layer models successfully capture the layer thickness $d_{(T)}$ through the interplay of cooling, enthalpy transport, and conduction \citep{tan21-lines, chen23}, but they implicitly assume sheet-like interfaces. Once the cold phase fragments, the temperature-dependent isosurface area $A_{(T)}$ becomes equally important. That missing geometric factor is why planar mixing layers and turbulent boxes, even with identical cooling, conduction, and turbulent driving, produce very different PDFs. We expect this geometric picture to be most applicable to quasi-isobaric, radiatively cooling multiphase flows in which temperature isosurfaces remain well-defined and the interface is at least moderately resolved.

This perspective sharpens several observational implications. In the ISM, it provides a natural route to the large thermally unstable H\,{\footnotesize I} fractions inferred observationally \citep{murray18, koley19}. The bottom panel of our \autoref{fig:ISM_mixing_layer_vs_turb_box} shows that 
naively applying planar turbulent mixing layers to estimate the abundance of intermediate temperature gas under-predicts the mass fraction of thermally unstable H\,{\footnotesize I} by $\sim 2$ orders of magnitude, highlighting the central role of geometry. In the CGM, it boosts the amount of gas near $T\sim10^5$--$10^6$ K relative to plane-parallel mixing-layer expectations, which is directly relevant to the large O\,{\footnotesize VI} reservoir \citep{werk14, tumlinson11, prochaska17, mcquinn18}. Under ICM conditions appropriate for jellyfish tails, our wind-tunnel runs likewise depart from the mixing-layer limit: the cold phase shatters into clumps embedded in a hot cocoon, and the resulting PDF resembles the turbulent-box case rather than the mixing-layer case. The observational consequences of this clump-in-cocoon geometry for the X-ray--H$\alpha$ surface-brightness ratio are developed in a companion study \citep{chen26}, where we show that mixing layer models yield surface-brightness ratio predictions that are more than 3 orders of magnitude lower than observations. Accounting for the distinct morphology of X-ray and H$\alpha$ emitting gas bridges this gap. This bears broader relevance to jellyfish tails, cluster filaments, and galactic winds \citep{sun22, olivares2025, Cecil2002, Strickland2002}.

Our main conclusions are as follows:
\begin{itemize}
\item Planar mixing layers and turbulent boxes with the same ${\rm Da}$, $\mathcal{M}_{\rm turb}$, cooling, and conduction produce markedly different temperature PDFs. The difference is not the local microphysics of the interface, but the temperature dependence of the isosurface area $A_{(T)}$: nearly constant for sheet-like mixing layers, but strongly rising and then flattening in turbulent boxes.

\item The turbulent-box PDF is governed by a clump-to-sheet transition. Weak driving produces onion-skin shells around isolated cold clumps, whereas stronger driving and higher temperature cause those shells to percolate into extended sheets. This transition is quantified by the rise of the largest-component area covering fraction $f_A$, the turnover of the Betti number $b_0$ (which tracks the number of isolated components), the sign change of the Euler characteristic $\chi \simeq b_0-b_1$ (where $\chi > 0$ and $\chi <0$ indicate clump and sheet-like geometry respectively), and the increase of the area per component $A/b_0$, all of which show that the interface becomes larger and more connected with stronger driving and at higher temperature. 

\item The $A_{(T)}d_{(T)}$ decomposition cleanly separates geometry from microphysics. The layer thickness $d_{(T)}$ is set mainly by cooling and conduction and varies comparatively little across the turbulent-box sequence, whereas $A_{(T)}$ drives most of the PDF evolution. This also explains why changing the normalization of the conduction coefficient $\kappa$ matters in clump-dominated flows but not in sheet-like mixing layers. In sheet-like mixing layers, changing the normalization of $\kappa$ mainly rescales the interface thickness d(T) without altering the nearly uniform sheet area A(T), whereas in clump-dominated flows it also smooths away small structures and changes the radii and relative areas of concentric shells, so the PDF itself changes.

\item The same geometric control reappears under CGM cooling and in wind-tunnel flows. Mixing-layer models are therefore accurate only in the sheet-geometry limit; once the cold phase fragments, they systematically under-predict intermediate-temperature gas, especially near the hot phase. This has direct implications for thermally unstable H\,{\footnotesize I} in the ISM, O\,{\footnotesize VI}-bearing gas in the CGM, and H$\alpha$--X-ray correlations in jellyfish tails.
\end{itemize}

A systematic survey of the $({\rm Da},\mathcal{M}_{\rm turb})$ plane and its mapping to observable phase fractions and line ratios will be presented elsewhere (Chen \& Oh 2026, in preparation). Natural extensions of this work include MHD simulations, anisotropic conduction and viscosity, cosmic rays and other non-thermal physics, non-equilibrium chemistry/cooling, and more realistic stratified or global simulations tied directly to observables.

\section*{Acknowledgements}

We thank the anonymous referee for insightful comments that improved the quality of this work. We acknowledge NSF grant AST240752 and HST grant AR-17860 for support. This work made considerable use of the Stampede3 supercomputer through allocation TG-PHY240194 from the Advanced Cyberinfrastructure Coordination Ecosystem: Services \& Support (ACCESS) program, which is supported by National Science Foundation grants \#2138259, \#2138286, \#2138307, \#2137603, and \#2138296.

\section*{Data Availability}

The data underlying this article will be shared upon reasonable request to the corresponding author.



\bibliographystyle{mnras}
\bibliography{reference, master_references}




\appendix

\section{Scale-Dependence of the Damk{\"o}hler number and Temperature PDFs}
\label{sec:scale_dependence_of_Da_and_temperature_PDF}

The \Da number is a scale-dependent parameter because $t_{\rm mix} \propto L/v \propto L^{2/3}$ and $t_{\rm cool}$ is independent of $L$. Throughout this work, we characterize each of our turbulent box and mixing layer simulations using a singular Da computed on the integral scale of turbulence, which is of order the simulation box size. The turbulent box simulations give rise to a collection of cold cloudlets, and one can in principle calculate cloud-scale Da by measuring the individual cloud sizes. In \autoref{sec:vary-parameters}, we stated that the integral-scale Da is preferred because there is no characteristic cloud scale, and the cold phase survival criterion is formulated on the seed cloud scale \citep{gronke22-turb}. Additionally, we demonstrate in this appendix that the turbulent-box temperature PDF becomes representative only on scales large enough to sample the interface network fairly.

\begin{figure*}
\centering
\includegraphics[width=\textwidth]{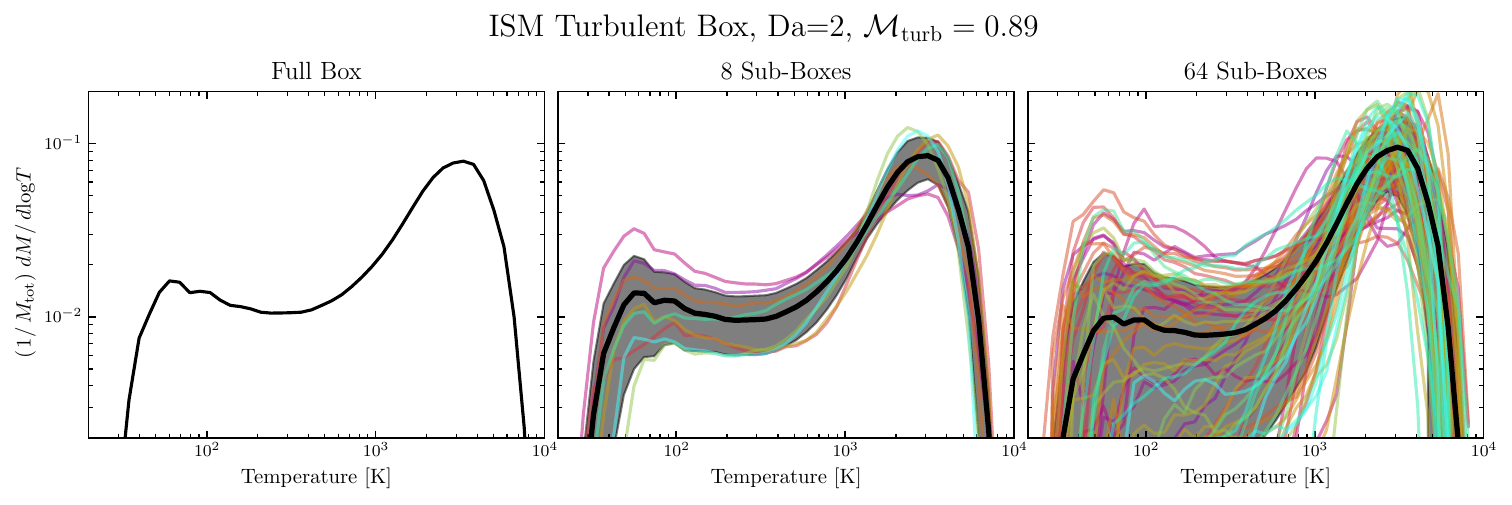}
\caption{Temperature PDFs from our Da=2, $\mathcal{M}_{\rm turb}=0.89$ ISM turbulent box simulation, including the full box PDF (\textit{left}) and PDFs from $2^3=8$ and $4^3=64$ equal-volume sub-boxes (\textit{middle} and \textit{right}). The sub-boxes are obtained by dividing the simulation domain evenly along each spatial dimension. Colored curves show PDFs of each individual sub-box, and the black curve shows the mean PDF, with the shaded region indicating the 1$\sigma$ scatter of the PDFs. The sub-box PDFs are reasonably similar to the full box PDF in the \textit{middle} panel but show large variations in the \textit{right} panel, where each sub-box has 1/4 the side length of the full box. For $T\gtrsim1000$K, the sub-box PDFs still look alike because higher temperature isosurfaces percolate into extended sheets and are well-represented in each sub-box. For $T\lesssim1000$K, geometric properties of lower temperature isosurfaces are not fairly sampled in different sub-boxes, resulting in very different PDF shapes. This investigation shows that a unique PDF only exists on scales approaching the box size. As such, we characterize our simulations using the box-scale Da instead of smaller cloud-scale Da.}
\label{fig:ISM_turb_box_Da_2_sub_box_PDFs}
\end{figure*}

In \autoref{fig:ISM_turb_box_Da_2_sub_box_PDFs}, we show temperature PDFs from our Da=2, $\mathcal{M}_{\rm turb}=0.89$ ISM turbulent box simulation, including the full box PDF (left panel) and PDFs measured in equal-volume sub-boxes. These sub-boxes are obtained by dividing the simulation domain evenly along each spatial dimension. Dividing each dimension in half results in $2^3=8$ sub-boxes (middle panel), and dividing each dimension into four equal segments results in $4^3=64$ sub-boxes (right panel). Colored curves show PDFs of each individual sub-box, and black curve shows the mean PDF, with the shaded region indicating the 1$\sigma$ scatter of the PDFs. In the middle panel where each PDF is measured on a scale of $L/2$, the individual PDFs are still reasonably similar to the full box PDF and only show modest variations across different sub-boxes, but on a scale of $L/4$ shown in the right panel, individual sub-box PDFs look completely different from each other. In particular, the sub-box PDFs are reasonable similar for $T\gtrsim1000$K but differ substantially for $T\lesssim1000$K. This is because the higher temperature isosurfaces percolate into extended sheets and are well-represented in each sub-box, while geometric properties of lower temperature isosurfaces are not fairly sampled in different sub-boxes. This means there is not a unique temperature PDF associated with a sub-box-scale Da. A representative PDF emerges only on scales large enough to sample the interface network fairly.

\begin{figure}
\centering
\includegraphics[width=\columnwidth]{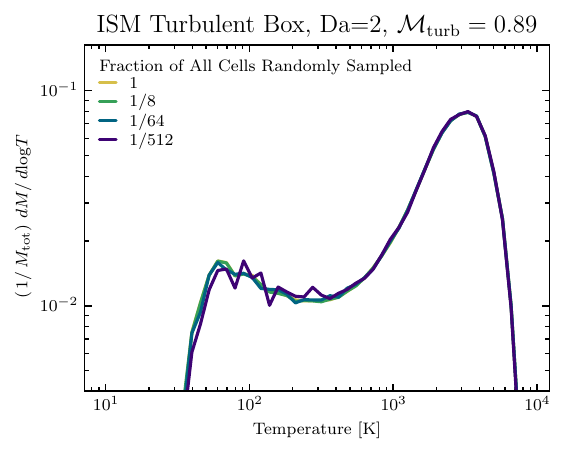}
\caption{Temperature PDFs from our Da=2, $\mathcal{M}_{\rm turb}=0.89$ ISM turbulent box simulation, obtained by randomly sampling a subset of the simulation cells. This is different from the sub-box approach in \autoref{fig:ISM_turb_box_Da_2_sub_box_PDFs} because the sampled subset of simulation cells are randomly distributed throughout the simulation domain. Unlike results in \autoref{fig:ISM_turb_box_Da_2_sub_box_PDFs}, temperature PDFs are well-converged with respect to the fraction of cells sampled. This demonstrates that the variations found in \autoref{fig:ISM_turb_box_Da_2_sub_box_PDFs} are not simply due to shot noise, but reflects an omission of some geometric features on the sub-box scale.}
\label{fig:ISM_turb_box_Da_2_PDF_random_sampling_cells}
\end{figure}

To demonstrate that the large variations in sub-box temperature PDFs shown in the right panel of \autoref{fig:ISM_turb_box_Da_2_sub_box_PDFs} are not simply due to shot noise, we compute temperature PDFs from the same ISM turbulent box by randomly sampling a subset of the simulation cells in \autoref{fig:ISM_turb_box_Da_2_PDF_random_sampling_cells}. Unlike the sub-box sampling shown in \autoref{fig:ISM_turb_box_Da_2_sub_box_PDFs}, the cells selected for computing PDFs shown in \autoref{fig:ISM_turb_box_Da_2_PDF_random_sampling_cells} are randomly distributed throughout the simulation domain and should fairly sample all the salient geometric features discussed in this work. Indeed, \autoref{fig:ISM_turb_box_Da_2_PDF_random_sampling_cells} shows that the temperature PDF is very well converged with respect to the fraction of cells randomly sampled. Randomly picking 1/64 of all simulation cells yields a temperature PDF identical to the full box PDF, while PDFs from sub-boxes with 1/64 the total volume vary significantly from each other and the full box PDF. This highlights the idea that sub-box-scale sampling omits key geometrical effects that set the temperature PDF.

\begin{figure*}
\centering
\includegraphics[width=\textwidth]{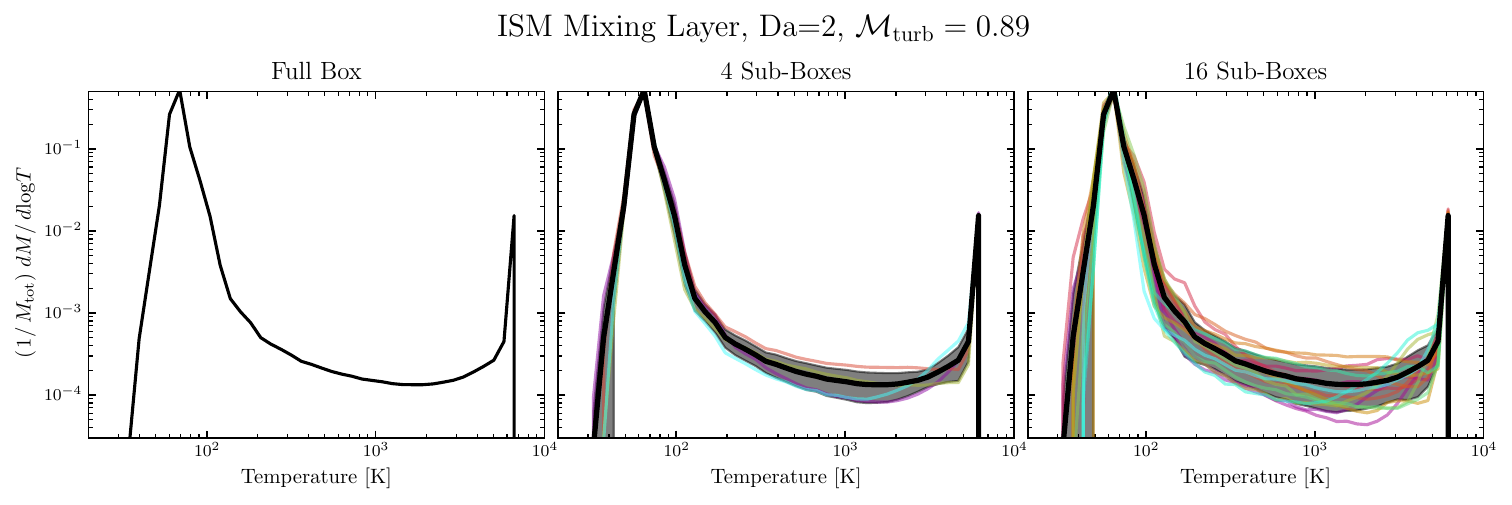}
\caption{Similar to \autoref{fig:ISM_turb_box_Da_2_sub_box_PDFs}, but for our ISM mixing layer simulation with Da=2, $\mathcal{M}_{\rm turb}=0.89$. When dividing the simulation domain into equal volume sub-boxes, we only partition the two dimensions parallel to the mixing layer interface so that all sub-boxes contain an equal fraction of the mixing interface. In a mixing layer setup, sub-box PDFs show much less variation compared to the full box PDF because the restrictive plane-parallel geometry forces all intermediate temperature isosurfaces to stack on top of each other as parallel sheets (\autoref{fig:ISM_mixing_layer_Da_2_temperature_contours}), and each sub-box samples a portion of this structure.}
\label{fig:ISM_mixing_layer_Da_2_sub_box_PDFs}
\end{figure*}

Mixing layer temperature PDFs do not suffer from variations due to sub-box sampling. \autoref{fig:ISM_mixing_layer_Da_2_sub_box_PDFs} is similar to \autoref{fig:ISM_turb_box_Da_2_sub_box_PDFs}, but for an ISM mixing layer simulation with Da=2 and $\mathcal{M}_{\rm turb}=0.89$. When dividing the simulation domain into equal volume sub-boxes, we only partition the two dimensions that are parallel to the mixing layer interface so that all sub-boxes contain an equal fraction of the mixing interface. \autoref{fig:ISM_mixing_layer_Da_2_sub_box_PDFs} shows that even when we divide each of these two dimensions into four equal segments, resulting in $4^2=16$ sub-boxes (right panel), the sub-box temperature PDFs are still consistent with each other and the full-box PDF. At this same $L/4$ scale, the turbulent box sub-box PDFs strongly discrepant with one another. The mixing layer setup does not suffer from this problem because its restrictive plane-parallel geometry forces all intermediate temperature isosurfaces to stack on top of each other as parallel sheets (\autoref{fig:ISM_mixing_layer_Da_2_temperature_contours}), and each sub-box samples a portion of this structure.


These experiments should not be interpreted as showing that the temperature PDF is a universal function of scale-dependent Da. Indeed, it shows the opposite: at fixed sub-box size, different contiguous regions can have very different PDFs because they sample different parts of the interface network. The PDF converges only when the sampled region is large enough to include a representative set of the relevant geometric structures. Thus, the failure of sub-box PDFs to converge is a geometric/intermittency effect.

\section{Convergence Test} \label{sec:convergence test}

\begin{figure}
\centering
\includegraphics[width=\columnwidth]{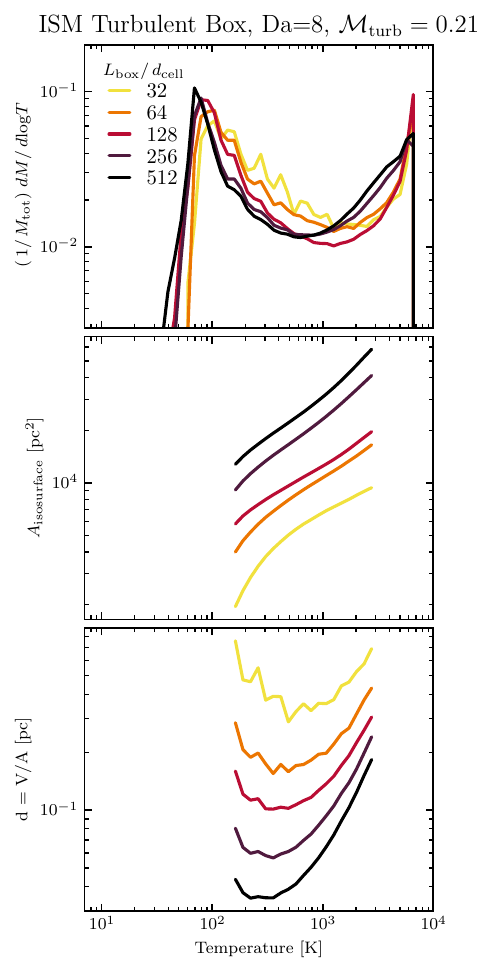}
\caption{\textit{Top}: Temperature PDF of our ISM turbulent box simulations is converged at a resolution of $\left. L_{\rm box} \right/ d_{\rm cell} = 256$. The fiducial resolution of the simulations we use in this work is $\left. L_{\rm box} \right/ d_{\rm cell} = 512$. \textit{Middle and Bottom}: Area of temperature isosurface and characteristic thickness obtained by decomposing the PDFs at different resolutions in the \textit{top} panel. The area of the fractal and corrugated temperature isosurfaces increase with resolution, and thickness decreases to keep the volume constant. The shapes of both $A_{(T)}$ vs. $T$ and $d_{(T)}$ vs. $T$ are numerically converged when the PDF is numerically converged.}
\label{fig:ISM_turb_box_convergence_test}
\end{figure}

We demonstrate the convergence of our numerical simulations by using the ISM turbulent box simulation with ${\rm Da} = \left. t_{\rm mix}\right/ t_{\rm cool} = 8$ and $\mathcal{M}_{\rm turb}=0.21$ as an example. Details of the simulation setup is described in \autoref{sec:methods}. In the top panel of \autoref{fig:ISM_turb_box_convergence_test}, we show mass-weighted temperature PDFs (which is the main simulation output being used throughout this work) from running the aforementioned simulation with resolution up to the fiducial choice of $512^3$ and down to $32^3$. The temperature PDFs are numerically converged at a resolution of $256^3$, which is below our fiducial resolution choice of $512^3$. In the middle and bottom panels of \autoref{fig:ISM_turb_box_convergence_test}, we decompose the temperature PDFs at a range of different resolutions into the area of temperature isosurfaces times the characteristic thickness of that temperature layer, as discussed in \autoref{sec:PDF-decomposition}. Since the temperature isosurfaces are fractal and corrugated, their surface areas increase with resolution. At the same time, the characteristic thickness decreases. Beyond a resolution of $256^3$, these two effects cancel out to ensure that the PDFs are numerically converged. The shapes of both $A_{(T)}$ vs. $T$ and $d_{(T)}$ vs. $T$ are numerically converged when the PDF is numerically converged.

As for the mixing layer simulations, we choose the parameters of our simulations such that the resolution per cell is identical to the fiducial turbulent box simulations described above. Thus, numerical convergence is expected to hold. For a detailed discussion on numerical convergence in mixing layer simulations, we refer the readers to \cite{tan21}.

\section{Dependence of Mixing Layer PDF on Box Height}
\label{sec:mixing_layer_PDF_vs_box_height}

\begin{figure}
\centering
\includegraphics[width=\columnwidth]{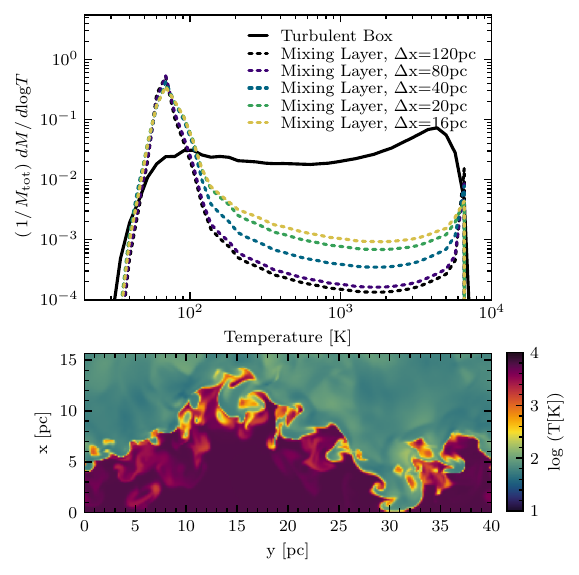}
\caption{\textit{Top}: Temperature PDFs of the ISM mixing layer simulation with Da$=2$, $\mathcal{M}_{\rm turb}=0.89$ (dotted lines), computed with varying height extents from the full simulation box ($\Delta$x=120pc) to $\sim 10\%$ the box ($\Delta$x=16pc) centered on the mixing layer interface. Although the relative abundance of intermediate temperature gas increases with decreasing height extent, the shape of the mixing layer PDFs in the intermediate temperature range remains unchanged and looks completely different from the turbulent box PDF (solid line). \textit{Bottom}: Temperature slice of the ISM mixing layer simulation discussed above with $\Delta$x=16pc, which is the smallest $\Delta$x choice in the \textit{top} panel. At $\Delta$x=16pc, the selected slab tightly encloses the active interface region, and further reducing $\Delta$x truncates features of the interface. The mixing layer PDF computed with this height extent still contains much less intermediate temperature gas compared to its turbulent box counterpart.}
\label{fig:mixing_layer_PDF_vs_box_height}
\end{figure}

When interpreting the mixing layer temperature PDFs presented throughout this work, it is worth noting that the amount of thermally stable gas (i.e. the height of the stable phase peaks in the PDFs) is set arbitrarily by the height of the simulation box. To understand how box height affects the mixing layer temperature PDF, we use the ISM mixing layer simulation with Da$=2$, $\mathcal{M}_{\rm turb}=0.89$ as an example and plot a series of temperature PDFs from the same simulation but computed with varying height extents in the top panel of \autoref{fig:mixing_layer_PDF_vs_box_height}. As we change the height extent from the full box ($\Delta$x=120pc) to $\sim 10\%$ the box ($\Delta$x=16pc) centered on the mixing layer interface, the amount of thermally stable gas decreases, which results in an upward shift of the intermediate temperature region of the PDF. Despite this shift in normalization, the shape of the mixing layer PDF in the intermediate temperature range remains unchanged and looks completely different from the turbulent box PDF. We discuss this difference in detail in \autoref{sec:turbulent box and mixing layer} and argue that it can be understood using a geometric decomposition of the PDF introduced in \autoref{sec:PDF-decomposition}. The bottom panel of \autoref{fig:mixing_layer_PDF_vs_box_height} shows a slice of our mixing layer simulation with $\Delta$x=16pc, which is the shortest height extent we chose in the top panel. At $\Delta$x=16pc, the selected slab tightly encloses the active interface region, and further reducing $\Delta$x truncates features of the interface. Even with $\Delta$x=16pc, the mixing layer PDF still contains much less intermediate temperature gas compared to its turbulent box counterpart, as shown in the top panel.

\section{Time Evolution of the Area-Thickness Decomposition}
\label{sec:time_evolution_of_A_d_decomposition}

\begin{figure}
\centering
\includegraphics[width=\columnwidth]{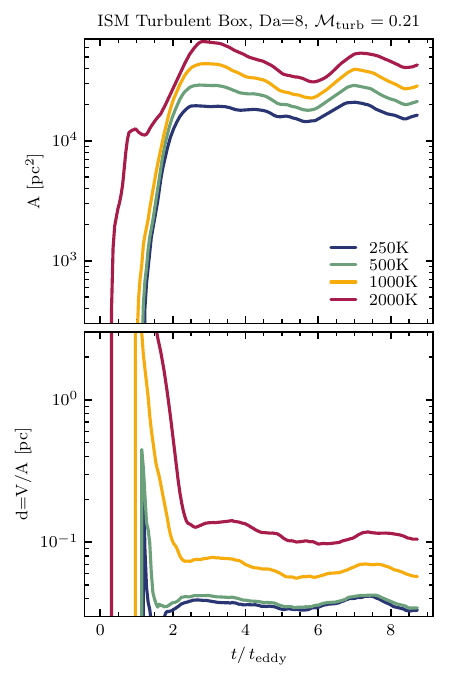}
\caption{Time evolution of the area and thickness of the T=250K, 500K, 1000K, and 2000K isosurfaces in the ISM turbulent box simulation with Da=8, $\mathcal{M}_{\rm turb}=0.21$. Both area and thickness are roughly time-independent once the temperature PDF reaches steady-state at $t \gtrsim 2.5 t_{\rm eddy}$. Mild fluctuations in both area and thickness over time occur concurrently across different temperatures so that the shapes of $A_{(T)}$ and $d_{(T)}$ remains unaffected.}
\label{fig:time_evolution_of_A_d_decomposition}
\end{figure}

All temperature PDFs presented in this work are time-independent, steady-state PDFs. However, it is useful to verify whether the area-thickness decomposition is also time-independent. We take the ISM turbulent box simulation with Da=8, $\mathcal{M}_{\rm turb}=0.21$ as an example and show how the area and thickness of the T=250K, 500K, 1000K, and 2000K isosurfaces evolve with time in \autoref{fig:time_evolution_of_A_d_decomposition}. This simulation starts from an initial condition of a uniform box at the unstable equilibrium temperature in the ISM, which means the isosurfaces are not well-defined in the first $t_{\rm eddy}$, the outer-scale eddy turnover time. At $t \sim 1-2.5 t_{\rm eddy}$, a multi-phase medium forms, and both area and thickness undergo rapid time evolution as the temperature PDF also evolves with time. Beyond $t \sim 2.5 t_{\rm eddy}$, the temperature PDF stabilizes and become time-independent, and both area and thickness become roughly time-independent as well. Although there are some mild fluctuations in both area and thickness over time, these fluctuations occur concurrently across different temperatures so that the shapes of $A_{(T)}$ and $d_{(T)}$ remain unaffected.

\end{document}